\documentclass[11pt,a4paper]{article}
\usepackage[margin=2.5cm]{geometry}
\usepackage{amsmath,amssymb,amsthm}
\usepackage{enumitem}
\usepackage{tikz}
\usetikzlibrary{arrows.meta,decorations.pathmorphing}
\usepackage[hidelinks]{hyperref}
\newtheorem{Theorem}{Theorem}
\newtheorem{Lemma}{Lemma}
\newtheorem{Proposition}{Proposition}
\newtheorem{Corollary}{Corollary}
\theoremstyle{definition}
\newtheorem{Definition}{Definition}
\numberwithin{equation}{section}
\newcommand{\X}{{\cal X}}
\newcommand{\Sset}{{\cal S}}
\newcommand{\E}{\mathbb{E}}
\newcommand{\bx}{\boldsymbol{x}}

\title{Universal Work Extraction from Individual Sequences\\ by Finite--State Information Engines}
\author{Neri Merhav\\[2mm]
\small The Andrew \& Erna Viterbi Faculty of Electrical and Computer Engineering\\
\small Technion---Israel Institute of Technology, Technion City, Haifa 3200003, Israel\\
\small E--mail: merhav@ee.technion.ac.il}
\date{}

\begin{document}\sloppy
\maketitle

\begin{abstract}
We consider information engines that interact with a heat bath at temperature $T$, a work reservoir, and a tape carrying an individual (deterministic) sequence $x$ over a finite alphabet $\X$. An engine is a cascade of autonomous ratchets with fixed energy levels and detailed--balance dynamics, each processing segments of $m$ tape cells and having finitely many internal states. We prove a converse bound on the extractable work in terms of the empirical block entropies of $x$, and show that it is achievable \emph{universally}, by engines whose energy levels do not depend on $x$. The maximal work per symbol is $kT[\ln|\X|-\bar H(x)]$, where $k$ is Boltzmann's constant and $\bar H(x)$ is the asymptotic normalized aligned block entropy of $x$, which equals its block Lempel--Ziv complexity, and the minimal work needed to erase $x$ is $kT\bar H(x)$ per symbol. Underlying these results is a two--way correspondence with universal coding: energy levels define a code, and a suitably operated cascade extracts $kT$ times $\ln|\X|$ per symbol minus the code length; conversely, a fluctuation identity shows that every engine fixed in advance defines a code. Hence, converses of universal coding apply verbatim to engines; e.g., the optimal redundancy of memoryless universal engines is given by the Shtarkov sum.
\end{abstract}


\section{Introduction}

In the model of Mandal and Jarzynski \cite{MJ12}, an autonomous device interacts sequentially with the bits of a tape, with a heat bath, and with a work reservoir, and it extracts work at the price of increasing the entropy of the tape. Many extensions followed, e.g., \cite{BS13,BMC16,BMC17,BMC18,DJ13}. In particular, Boyd, Mandal and Crutchfield \cite{BMC16,BMC17} considered general finite--state ratchets driven by correlated random inputs, and in \cite{p193}, work/entropy--production inequalities were derived for general finite--state ratchets and general (possibly non--stationary) random inputs. In all these works, the tape is random and the informational ingredient of the second law is a Shannon entropy.

In \cite{p180}, the question was raised what replaces the input entropy when the tape carries an \emph{individual sequence} $x=(x_1,x_2,\ldots)$, i.e., a deterministic sequence, over a finite alphabet $\X$, with no underlying probabilistic model. For a memoryless device that interacts with one bit at a time, it was shown that the Lempel--Ziv (LZ) complexity \cite{ZL78} of the input plays this role. A finite--state restriction is unavoidable here: with a deterministic input, the bound of \cite{p193} does not depend on the input at all, and a device with enough memory to store the first $N$ symbols, $x^N=(x_1,\ldots,x_N)$, is, in principle, not limited by their structure. The structure of $x$ can matter only if the complexity of the device is bounded independently of $N$, which is the finite--state paradigm of Ziv and Lempel \cite{ZL78}.

In this paper, we consider engines that are cascades of autonomous ratchets with fixed energy levels, each ratchet processing non--overlapping segments of $m$ tape cells and having finitely many internal states; the product of the numbers of internal states of the ratchets is at most $s$. The heat bath is at temperature $T$, and $k$ denotes Boltzmann's constant. Our main results are the following.
\begin{enumerate}[label=(\roman*),leftmargin=*]
\item A converse bound (Theorem~\ref{thm:conv}): the work per symbol is at most $kT[\ln|\X|-\hat H_\ell(x^N)/\ell]$, where $\hat H_\ell(x^N)$ is the empirical entropy of the non--overlapping blocks of $\ell$ consecutive symbols of $x^N$, for every $\ell$ that is a multiple of $m$, up to a penalty for the memory of the engine that vanishes as $\ell$ grows.
\item Universal achievability (Theorem~\ref{thm:ach}): a cascade of memoryless ratchets, whose energy levels are fixed in advance, attains this bound for every $x$, up to the redundancy of a universal code. The mechanism is simple: the cascade, operated as a quasi--static heating in space, extracts $kT$ times $\ln|\X|$ per symbol minus a code length, and the code is determined by the energy levels.
\item The work capacity of $x$ is $kT[\ln|\X|-\bar H(x)]$, where $\bar H(x)$ is the asymptotic value of $\hat H_\ell(x^N)/\ell$ as $N\to\infty$ and then $\ell\to\infty$ (more precisely, the infimum over $\ell$ of the $\limsup_N$; see Theorem~\ref{thm:cap}). It is attained by a single sequence of engines that do not depend on $x$ (Theorem~\ref{thm:univ}), for example, engines whose energy levels are LZ78 code lengths (Corollary~\ref{cor:lzengine}). This characterizes the supremum over all segment lengths and memory sizes; the work capacity for a fixed segment length and a fixed number of internal states remains open. Moreover, $\bar H(x)$ equals the block LZ complexity of $x$, i.e., the asymptotic LZ complexity when the incremental parsing is restarted at the beginning of every block of a given length, which then tends to infinity (Proposition~\ref{prop:lz}); this is shown partly by a thermodynamic argument, together with the existence of the limit of the block LZ complexity, established in \cite{p246}.
\item A fluctuation identity yields a Kraft inequality for work: every memoryless engine fixed in advance defines a lossless code whose length is $\ln|\X|$ per symbol minus the work (Lemma~\ref{lem:kraftwork}), and for engines with memory, the same holds for the whole input, up to an additive $\ln s$ (Lemma~\ref{lem:kraftmem}). Consequently, the minimax redundancy of memoryless universal engines, relative to the target $\ln|\X|-\hat H_p(x^N)/p$, where $m=qp$ for positive integers $p$ and $q$, is exactly the minimax regret of universal coding for $q$ symbols from an alphabet of size $A=|\X|^p$ relative to the memoryless sources, i.e., $[\ln S_q(A)]/m$ per symbol, where $S_q(A)$ is the Shtarkov sum; it is attained, up to terms that vanish as the number of stages grows, by engines whose energy levels are proportional to the empirical entropy of the segment (Theorem~\ref{thm:redund}), and, by the individual--sequence lower bound of \cite{WMF94}, it cannot be avoided on most periodic inputs (Corollary~\ref{cor:most}).
\item Sequences that are incompressible by LZ78 are useless as fuel for every finite--state engine (Theorem~\ref{thm:incomp}). Dually, as a deterministic analogue of Landauer's principle, the minimal work needed to erase $x$ is $kT\bar H(x)$ per symbol, again attainable universally (Theorem~\ref{thm:erase}); hence the fuel value and the erasure cost of $x$ add up to $kT\ln|\X|$ per symbol.
\end{enumerate}
A theme that runs through these results is a two--way correspondence between universal coding and the thermodynamics of information engines. In one direction, codes yield engines: an energy landscape is a probability assignment, and a cascade operated as a quasi--static heating in space extracts $kT$ times $\ln|\X|$ per symbol minus the code length (Lemma~\ref{lem:landscape}), so that every universal code (type--class codes, the LZ78 code, the normalized maximum--likelihood code) yields a universal engine. In the other direction, engines yield codes: by a fluctuation identity, $\ln|\X|$ per symbol minus the work of any memoryless engine fixed in advance satisfies the Kraft inequality (Lemma~\ref{lem:kraftwork}), and the same holds, up to an additive $\ln s$, for engines with memory (Lemma~\ref{lem:kraftmem}), so that the converse theorems of universal coding, such as Shtarkov's minimax regret bound \cite{Shtarkov87} and the individual--sequence lower bound of \cite{WMF94}, apply verbatim to engines. Thus, the theory of universal coding of individual sequences provides both the constructions and the limitations of information engines.

The remaining part of the paper is organized as follows. In Section~\ref{sec:prelim}, we establish notation and recall the relevant complexity measures. In Section~\ref{sec:model}, we describe the system model. In Section~\ref{sec:conv}, we derive the converse bound, and in Section~\ref{sec:ach}, the universal achievability result. In Section~\ref{sec:cap}, we combine them into the characterization of the work capacity. In Section~\ref{sec:redund}, we derive the Kraft inequality for work and the optimal redundancy of memoryless universal engines, and in Section~\ref{sec:erase} we treat incompressible sequences and erasure. Section~\ref{sec:disc} concludes with a discussion, including numerical illustrations. Technical lemmas are proved in Appendices~\ref{app:blocks}--\ref{app:kraftmem}.

\subsection{Related work}

\paragraph{Information ratchets.} The model of Mandal and Jarzynski \cite{MJ12} and its extensions \cite{BS13,DJ13,BMC16,BMC17,BMC18,p193} consider random tapes, and the informational ingredient of their second laws is a Shannon entropy (rate). Boyd, Mandal and Crutchfield \cite{BMC17} showed that, to harvest the structure of a correlated random input, the memory of a ratchet should reflect the memory of the input process. In contrast, here the tape is an individual sequence, the informational ingredient is an empirical (finite--state) complexity, and the memory of the engine appears in the converse only through a penalty term that vanishes as the block length of the analysis grows; in the achievability part, the structure of the sequence is exploited through long segments, with no memory at all. Garner, Thompson, Vedral and Gu \cite{GTVG17} studied pattern manipulators, devices that convert work into patterns (temporally ordered data) or vice versa, and showed that their minimal dissipation is governed by complexity--theoretic properties of the (random) pattern; here the pattern is a single deterministic sequence, and the relevant complexity is its finite--state (LZ) complexity.

\paragraph{Work value of individual strings.} Bennett \cite{Bennett82} and Zurek \cite{Zurek89} related the work value and the erasure cost of an individual string to its algorithmic (Kolmogorov) complexity, for demons that are universal computers; see also \cite{BW19}. Kolchinsky and Wolpert \cite{KW20} analyzed physical realizations of Turing machines, related their heat generation to algorithmic complexity, and examined the erasure of a long string. Kolmogorov complexity is not computable, and the bounds involve unspecified additive constants. Here the demon is an autonomous thermal machine with finitely many internal states, the complexity measure is computable (empirical entropies and LZ complexity), and the converse bounds are explicit for every finite $N$. In \cite{p180}, the LZ complexity of the input was shown to bound the work extracted by a memoryless two--level device that interacts with one bit at a time. The present paper allows memory, segments of arbitrary length and cascades, proves matching achievability with engines fixed in advance, and identifies the capacity exactly.

\paragraph{Work extraction and universal coding.} The idea that optimal work extraction is related to universal data compression is not new. Touzo, Marsili, Merhav and Rold\'an \cite{TMMR20} considered Szil\'ard--type engines with measurement and feedback, showed that the optimal non--equilibrium state before the measurement is the normalized maximum--likelihood distribution of a statistical model, and related the maximal extracted work to the minimax redundancy of universal coding for that model. In Section~\ref{sec:redund}, a relation of a similar spirit is established for autonomous engines and individual sequences: the minimax redundancy of memoryless engines fixed in advance equals the minimax regret (the logarithm of the Shtarkov sum) of universal coding. Boyd, Crutchfield and Gu \cite{BCG22} showed that maximizing the work extracted by an agent designed from data is equivalent to maximum--likelihood estimation of a model of the data. Our setting differs in three respects: the tape is an individual sequence rather than a random one; the engines are autonomous ratchets with fixed energy levels, operated without measurement, feedback or driving; and the engine is fixed in advance rather than estimated from the data. The correspondence underlying our achievability results is that the energy landscape of the first stage plays the role of a probability assignment, so that the work equals $kT$ times $\ln|\X|$ per symbol minus a code length (Lemma~\ref{lem:landscape}), and universal codes (based on types, or LZ78) yield universal engines.

\paragraph{Mismatch cost.} Kolchinsky and Wolpert \cite{KW17} showed that when a fixed physical process is run on an initial distribution different from the one that minimizes its dissipation, the extra dissipation is given by a contraction of relative entropy, which they termed the mismatch cost. The loss of the first stage of our cascade, the divergence between the empirical distribution of the segments and the first energy landscape, is an instance of this cost. Our universal engines may be viewed as processes whose mismatch cost, per symbol, vanishes for every individual sequence, owing to the use of a universal probability assignment as the energy landscape.

\paragraph{Universal work extraction in quantum thermodynamics.} Watanabe and Takagi \cite{WT26} constructed a quantum work extraction protocol whose description does not depend on the input state, and which nevertheless attains the free energy of an unknown input state, given many independent copies. This is the same notion of universality, in a probabilistic (i.i.d.) quantum setting with general thermal operations. Here the input is a single deterministic sequence, and the engines are classical, autonomous and finite--state.

\section{Notation and Preliminaries}\label{sec:prelim}

Random variables are denoted by capital letters, their realizations by the corresponding lower case letters, and their alphabets by calligraphic letters. Throughout, $x=(x_1,x_2,\ldots)$ is an infinite deterministic sequence over a finite alphabet $\X$ of size $|\X|\ge2$, $x^N=(x_1,\ldots,x_N)$, and $x_i^j=(x_i,\ldots,x_j)$. All logarithms are natural logarithms, denoted by $\ln(\cdot)$, and accordingly, all entropies and information quantities are in nats. The entropy of a distribution $P$ on a finite set is $H(P)=-\sum_aP(a)\ln P(a)$. Following the custom in information theory, we also write $H(U)$ for the entropy of a random variable $U$, and $H(U|V)$ for conditional entropy \cite{CT06}. This is an abuse of notation, since $H(U)$ is not a function of the random outcome of $U$ but of its distribution, denoted $P_U$: $H(U)=H(P_U)$. The two usages are distinguished by the argument: $P$, $Q$, $R$, $\Pi$, $\mu$ (possibly with indices) always denote distributions (or, for $P$ with two arguments, transition matrices), and all other capital letters ($U$, $V$, $S$, $Y$, $I$, \ldots) denote random variables, so $H(P)$ is the entropy of the distribution $P$, and $H(U)$ that of the distribution of $U$. Similarly, $\hat H_\ell(x^N)$ and $\hat H_{\rm csw}(X^d)$ below are empirical entropies, i.e., entropies of empirical distributions computed from $x^N$. These are entropies of blocks, not normalized per symbol. A bar indicates a quantity normalized per symbol and taken in the limit $N\to\infty$, e.g., $\bar H_\ell(x)=\limsup_N\hat H_\ell(x^N)/\ell$ in \eqref{eq:rho} below, as opposed to $\hat H_\ell(x^N)$ itself. $D(P\|Q)$ is the divergence, $h(\epsilon)=-\epsilon\ln\epsilon-(1-\epsilon)\ln(1-\epsilon)$ is the binary entropy function, and ${\cal I}\{\cdot\}$ is the indicator function. $\E\{\cdot\}$ denotes expectation, and $\langle\cdot\rangle$ denotes expectation with respect to the thermal randomness of the dynamics.

\paragraph{Aligned empirical block entropies.} For a positive integer $\ell$, and $N$ an integral multiple of $\ell$, let $n=N/\ell$, and let $\bx_i=x_{i\ell+1}^{i\ell+\ell}$, $i=0,1,\ldots,n-1$, be the $n$ aligned $\ell$--blocks of $x^N$. Their empirical distribution and entropy are
\begin{equation}
\hat P_\ell(a)=\frac1n\sum_{i=0}^{n-1}{\cal I}\{\bx_i=a\},\quad a\in\X^\ell,\qquad \hat H_\ell(x^N)=H(\hat P_\ell).
\end{equation}
An elementary but important observation is the following: if $I$ is uniformly distributed over $\{0,\ldots,n-1\}$, then the random block $\bx_I$ has distribution $\hat P_\ell$, and so
\begin{equation}\label{eq:emp}
H(\bx_I)=\hat H_\ell(x^N).
\end{equation}
This allows ordinary information--theoretic inequalities to be applied to empirical quantities. For an infinite sequence $x$, let
\begin{equation}\label{eq:rho}
\bar H_\ell(x)=\limsup_{N\to\infty}\frac{\hat H_\ell(x^N)}{\ell},\qquad \bar H(x)=\inf_{\ell\ge1}\bar H_\ell(x),
\end{equation}
where $N$ ranges over the multiples of $\ell$ (whenever $\hat H_\ell(x^N)$ appears, $N$ is understood to be a multiple of $\ell$). Although $\bar H_\ell(x)$ need not be monotone in $\ell$, it is non--increasing along multiples: $\bar H_{k\ell}(x)\le \bar H_\ell(x)$ for every positive integer $k$ (Lemma~\ref{lem:mult} in Appendix~\ref{app:blocks}). Consequently, $\bar H(x)$ is also a limit: since every $\ell$ divides $k!$ for $k\ge\ell$, we have $\bar H(x)\le \bar H_{k!}(x)\le \bar H_\ell(x)$ for $k\ge\ell$, and so $\bar H(x)=\lim_{k\to\infty}\bar H_{k!}(x)$.

\paragraph{LZ complexity.} The incremental parsing of Lempel and Ziv \cite{ZL78} (LZ78) divides $x^N$ sequentially into phrases, each being the shortest string that has not yet appeared as a phrase; only the last phrase may be a repetition of an earlier one. For example, the binary string $010010110$ is parsed as
\begin{equation}
0\,|\,1\,|\,00\,|\,10\,|\,11\,|\,0 .
\end{equation}
Each new phrase consists of an earlier phrase (or the empty string) followed by one symbol. It can therefore be encoded by a pointer to that earlier phrase and the new symbol, and so the whole string is encoded with about $c\ln c$ nats, where $c=c(x^N)$ is the number of phrases (see Section~\ref{sec:lzland} for the exact code length). Accordingly, the \emph{LZ complexity}
\begin{equation}
\rho_{\rm LZ}(x^N)=\frac{c(x^N)\ln c(x^N)}{N}
\end{equation}
is, asymptotically, the compression ratio of LZ78 in nats per symbol. A sequence with much repetitive structure is parsed into few long phrases, and has small LZ complexity. A sequence with no structure has $c(x^N)\approx (N\ln|\X|)/\ln N$, and its LZ complexity is close to the maximal value, $\ln|\X|$.

The fundamental property of LZ78 is that, for every individual sequence, its compression ratio is asymptotically no larger than that of the best finite--state encoder \cite{ZL78}. A quantitative form of this property, in terms of aligned block entropies, is the following \cite[eq.~(21)]{Merhav13} (rewritten here in natural logarithms, with $\alpha=|\X|$): for every $\ell$ and every $N$ that is a multiple of $\ell$,
\begin{equation}\label{eq:H-LZ}
\frac{\hat H_\ell(x^N)}{\ell}\ge\rho_{\rm LZ}(x^N)-\delta(N,\ell),\qquad \delta(N,\ell)=\frac{2\ell(\ln\alpha+\ln2)^2}{(1-\epsilon_N)\ln N}+\frac{2\ell\alpha^{2\ell}\ln\alpha}{N}+\frac{\ln2}{\ell},
\end{equation}
where $\epsilon_N\to0$ (for $\alpha=2$, this is \cite[eq.~(37)]{p180}). As $N\to\infty$, $\delta(N,\ell)\to(\ln2)/\ell$, and so $\limsup_N\rho_{\rm LZ}(x^N)\le \bar H_\ell(x)+(\ln2)/\ell$ for every $\ell$. Applying this with $k\ell$ in place of $\ell$, using $\bar H_{k\ell}(x)\le \bar H_\ell(x)$, and letting $k\to\infty$, gives $\limsup_N\rho_{\rm LZ}(x^N)\le \bar H_\ell(x)$ for every $\ell$, that is, $\bar H(x)\ge\limsup_N\rho_{\rm LZ}(x^N)$, where now $N$ ranges over all positive integers. (Restricting $N$ to the multiples of $\ell$ is immaterial here: if $N'=\ell\lfloor N/\ell\rfloor$, then $c(x^{N'})\le c(x^N)\le c(x^{N'})+\ell$, since each additional symbol adds at most one phrase, and so $|\rho_{\rm LZ}(x^N)-\rho_{\rm LZ}(x^{N'})|=O[(\ell\ln N)/N]$.) In words: the LZ complexity cannot exceed the normalized block entropy by more than a vanishing amount.

\paragraph{Block LZ complexity.} The reverse inequality does not hold in general for the LZ complexity of the whole sequence. It does hold if the parsing is restarted at the beginning of every block of $m$ symbols, and $m$ grows \cite{p246}. Specifically, for a positive integer $m$, let
\begin{equation}
\rho_m(x)=\limsup_{N\to\infty}\frac mN\sum_{i=0}^{N/m-1}\rho_{\rm LZ}(x_{im+1}^{im+m}),
\end{equation}
with $N$ ranging over multiples of $m$; this is the compression ratio of LZ78 applied separately to consecutive blocks of length $m$. It is the natural LZ quantity for engines that process segments of length $m$ (Section~\ref{sec:lzland}). In \cite{p246}, $\rho_m(x)$ is compared with \emph{cyclic sliding--window} empirical entropies, which, unlike aligned block entropies, count the $d$--tuples starting at every position of $x^N$. Specifically, for a positive integer $d$, the cyclic sliding--window empirical distribution of $d$--tuples is
\begin{equation}
\hat P_{\rm csw}(a)=\frac1N\sum_{i=1}^{N}{\cal I}\{x_i^{i+d-1}=a\},\qquad a\in\X^d,
\end{equation}
where indices are taken modulo $N$, i.e., $x_{N+j}=x_j$ for $j=1,\ldots,d-1$ (the wrap--around makes the empirical distributions of different orders consistent with each other: the marginal of $\hat P_{\rm csw}$ on the first $d'<d$ coordinates is the cyclic sliding--window empirical distribution of $d'$--tuples). Following \cite{p246}, its entropy is denoted by $\hat H_{\rm csw}(X^d)=H(\hat P_{\rm csw})$, and we let
\begin{equation}
\bar H_{\rm csw}[x]=\lim_{d\to\infty}\limsup_{N\to\infty}\frac{\hat H_{\rm csw}(X^d)}d,
\end{equation}
where the limit exists by subadditivity. The main subject of \cite{p246} is chain rules for LZ complexity; along the way, matching upper and lower bounds on $\rho_m(x)$ in terms of cyclic sliding--window entropies are derived \cite[eqs.~(24), (37)]{p246}, and, as a byproduct, \cite[eq.~(38)]{p246}
\begin{equation}\label{eq:p246}
\lim_{m\to\infty}\rho_m(x)=\bar H_{\rm csw}[x].
\end{equation}
(The definitions above are those of \cite[eqs.~(9)--(13), (18)--(21)]{p246}; there, logarithms are to base 2, which multiplies both sides of \eqref{eq:p246} by the same factor.) In Section~\ref{sec:cap}, we show that $\lim_m\rho_m(x)=\bar H(x)$; thus aligned block entropies, cyclic sliding--window entropies, and block LZ complexity all lead to the same quantity $\bar H(x)$.

\section{System Model}\label{sec:model}

\paragraph{A single ratchet.} As in \cite{MJ12,p180,p193}, the system consists of a heat bath at temperature $T$, a work reservoir, an input tape carrying $x$, an output tape, and a device (ratchet). The tape is divided into consecutive, non--overlapping segments of $m$ cells, which the ratchet processes one per cycle. The ratchet has a finite set $\Sset$ of $s$ internal states ($s=1$ means that it is memoryless).

There are therefore two different notions of a state, which should not be confused. A \emph{ratchet state} $\sigma\in\Sset$ is the internal state of the device, which it carries from one cycle to the next and which constitutes its memory. A \emph{Markov state} $(\xi,\sigma)\in\X^m\times\Sset$ is the pair consisting of the content $\xi$ of the segment currently held by the ratchet and the ratchet state $\sigma$; it is the state of the Markov jump process that describes the dynamics within a cycle. Energies and transition rates are attached to Markov states. Whenever there is room for ambiguity, we use the terms ``ratchet state'' and ``Markov state'' explicitly.

\begin{figure}[ht]
\centering
\begin{tikzpicture}[>=Stealth, font=\small, scale=0.97, transform shape,
  cell/.style={draw, minimum width=5mm, minimum height=5mm, inner sep=0pt},
  box/.style={draw, thick, rounded corners, minimum width=26mm, minimum height=16mm, align=center}]
\node[draw, thick, minimum width=44mm, minimum height=8mm, fill=gray!10] (bath) at (0,2.6) {heat bath, temperature $T$};
\node[box] (r) at (0,0) {ratchet\\[1mm] state $\sigma\in\Sset$};
\draw[->, thick, decorate, decoration={snake, amplitude=0.6mm, segment length=3mm}] (bath.south) -- node[right]{heat} (r.north);
\foreach \i/\v in {0/0,1/1,2/1,3/0,4/1,5/0} {\node[cell] (in\i) at (-6.2+0.5*\i,0) {\v};}
\draw[thick] (-6.45,-0.25) rectangle (-5.45,0.25);
\draw[thick] (-5.45,-0.25) rectangle (-4.45,0.25);
\draw[thick] (-4.45,-0.25) rectangle (-3.45,0.25);
\node at (-4.95,-0.6) {input tape ($x$), segments of $m$ cells};
\draw[->, thick] (-3.35,0) -- node[above]{$U_i$} (r.west);
\foreach \i/\v in {0/1,1/0,2/0,3/1,4/1,5/1} {\node[cell] (out\i) at (3.7+0.5*\i,0) {\v};}
\draw[thick] (3.45,-0.25) rectangle (4.45,0.25);
\draw[thick] (4.45,-0.25) rectangle (5.45,0.25);
\draw[thick] (5.45,-0.25) rectangle (6.45,0.25);
\node at (4.95,-0.6) {output tape};
\draw[->, thick] (r.east) -- node[above]{$V_i$} (3.35,0);
\node[draw, minimum width=12mm, minimum height=6mm] (d) at (0,-2.0) {delay};
\draw[->, thick] ([xshift=10mm]r.south) -- ++(0,-0.95) node[midway, right]{$S_{i+1}$} |- (d.east);
\draw[->, thick] (d.west) -| ([xshift=-10mm]r.south) node[pos=0.75, left]{$S_i$};
\draw[thick] (2.6,2.0) circle (3.5mm);
\draw[thick] (2.95,2.0) -- (2.95,1.2);
\node[draw, thick, fill=gray!30, minimum width=5mm, minimum height=4mm] at (2.95,1.0) {};
\node[align=left, anchor=west] at (3.3,1.4) {work reservoir};
\draw[->, thick] (r.north east) -- node[above left=-0.5mm]{work} (2.35,1.8);
\end{tikzpicture}

\vspace{4mm}
\begin{tikzpicture}[>=Stealth, font=\small,
  st/.style={draw, thick, rounded corners, minimum width=15mm, minimum height=8mm, align=center}]
\node (x) at (0,0) {$x$};
\node[st] (s1) at (1.8,0) {stage 1\\[-0.5mm]\footnotesize $s_1$ states};
\node[st] (s2) at (4.3,0) {stage 2\\[-0.5mm]\footnotesize $s_2$ states};
\node (dots) at (6.3,0) {$\cdots$};
\node[st] (sL) at (8.3,0) {stage $L$\\[-0.5mm]\footnotesize $s_L$ states};
\node (y) at (10.3,0) {output};
\draw[->, thick] (x) -- (s1); \draw[->, thick] (s1) -- node[above]{\footnotesize tape 1} (s2);
\draw[->, thick] (s2) -- (dots); \draw[->, thick] (dots) -- (sL); \draw[->, thick] (sL) -- (y);
\end{tikzpicture}
\caption{Top: a single ratchet. In cycle $i$, the ratchet interacts with the $i$--th segment $U_i$ of $m$ tape cells, with the heat bath and with the work reservoir, during an interaction interval of length $\tau$; it then writes the outgoing segment $V_i$ and passes its new ratchet state $S_{i+1}$ to the next cycle. Bottom: an engine is a cascade of $L$ such ratchets (stages), each processing the output tape of the previous one, on the same grid of segments.}
\label{fig:model}
\end{figure}
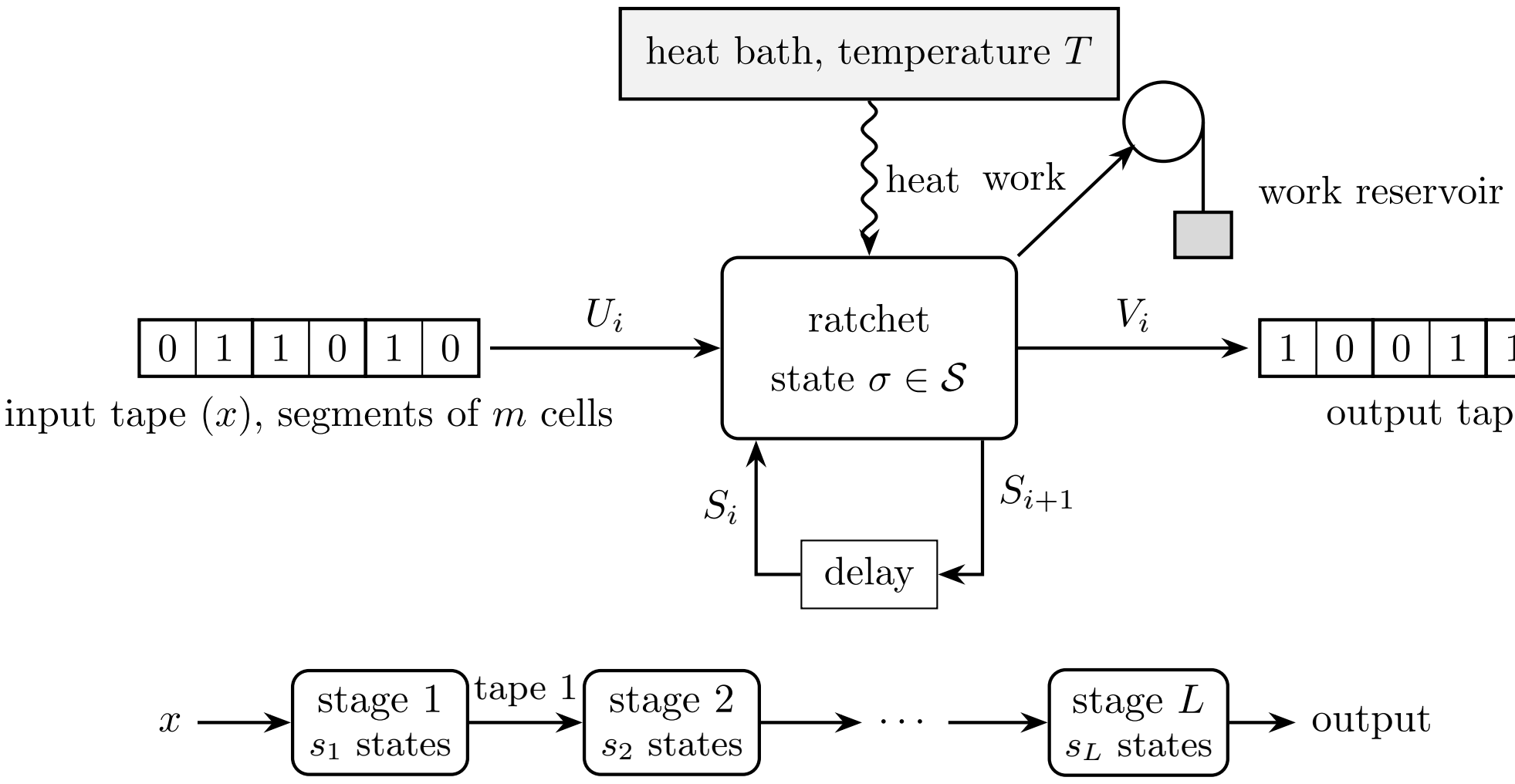

In cycle $i$, the Markov state $(\xi_t,\sigma_t)\in\X^m\times\Sset$ starts at $(U_i,S_i)$, where $U_i$ is the $i$--th incoming segment and $S_i$ is the current ratchet state, and it evolves during an interaction interval of length $\tau$ as a Markov jump process with transition rates $M[(\xi,\sigma)\to(\xi',\sigma')]$. At the end of the interval, $V_i=\xi_\tau$ is recorded on the output tape and $S_{i+1}=\sigma_\tau$ becomes the next ratchet state. Each Markov state has a fixed energy $E(\xi,\sigma)$, and the rates obey detailed balance with respect to the canonical distribution
\begin{equation}\label{eq:canonical}
\Pi(\xi,\sigma)=\frac{e^{-\beta E(\xi,\sigma)}}{Z(\beta)},\qquad \beta=\frac1{kT},
\end{equation}
where $T$ is the temperature of the heat bath, $k$ is Boltzmann's constant, and $Z(\beta)=\sum_{\xi,\sigma}e^{-\beta E(\xi,\sigma)}$ is the partition function.
As in \cite{p193}, each jump $(\xi,\sigma)\to(\xi',\sigma')$ lifts the mass of the work reservoir and thereby performs the work $E(\xi',\sigma')-E(\xi,\sigma)$, extracted from the heat bath, whereas the switch from one cycle to the next is not accompanied by any exchange of energy. Hence the work of cycle $i$ is $E(V_i,S_{i+1})-E(U_i,S_i)$. The quantities $U_i,S_i,V_i$ are random because of the thermal fluctuations; for the first ratchet, $U_i$ is a deterministic segment of $x$. For $m=1$, this is the model of \cite{p193}, and for $m=s=1$, that of \cite{p180}. For $m>1$, the energy levels $E(\xi,\sigma)$ may couple all $m$ cells of a segment, and this coupling is the means by which the engines of Section~\ref{sec:ach} harvest the structure of $x$; the converse bound of Section~\ref{sec:conv} holds for the same model, so the two are directly comparable. The model is depicted in Figure~\ref{fig:model}.

\paragraph{Engines.} An engine is a cascade of $L$ ratchets (stages) with a common segment length $m$ and $s_1,\ldots,s_L$ internal states. Stage 1 processes the input tape, and stage $r\ge2$ processes the output tape of stage $r-1$, on the same grid of segments. Since a cycle of stage $r$ acts only on one segment of tape $r-1$ and on the ratchet state of stage $r$, cycles acting on disjoint variables commute, and we may regard the stages as operating one after another. $W_N$ denotes the total work of all stages on the first $N$ input symbols ($N$ a multiple of $m$).

\paragraph{Realizing a ratchet with prescribed energy levels.} The model leaves the transition rates free, subject only to detailed balance. Here is a standard way to construct them, given any prescribed energy levels $E(z)$ of the Markov states $z=(\xi,\sigma)\in\X^m\times\Sset$. As in \cite{p193}, $E(z)$ may be thought of as the height of the load of the work reservoir when the system is in Markov state $z$, so that each jump lifts or lowers the load by the corresponding energy difference.
\begin{itemize}
\item \emph{Allowed transitions.} Choose a connected graph $G$ on $\X^m\times\Sset$ whose edges are the allowed transitions; for example, $z$ and $z'$ are neighbors if they differ in a single cell of the segment, or in the ratchet state only (single--flip moves), or $G$ is the complete graph.
\item \emph{Rates.} For every edge $\{z,z'\}$ of $G$, choose a symmetric barrier $B(z,z')=B(z',z)$, with $B(z,z')\ge\max\{E(z),E(z')\}$, and let
\begin{equation}\label{eq:arrhenius}
M[z\to z']=\begin{cases}\nu\exp\bigl\{-\beta[B(z,z')-E(z)]\bigr\}, & \{z,z'\}\in G,\\ 0, & \text{otherwise},\end{cases}
\end{equation}
where $\nu>0$ sets the time scale. This is the Arrhenius (transition--state) form: a jump requires thermal activation over the barrier separating the two states.
\end{itemize}
Then $\Pi(z)M[z\to z']=[\nu/Z(\beta)]e^{-\beta B(z,z')}$ is symmetric in $z$ and $z'$, which is exactly detailed balance with respect to \eqref{eq:canonical}; and since $G$ is connected, the process is irreducible (every Markov state can be reached from every other through jumps of positive rate; see Appendix~\ref{app:markov}), so $\Pi$ is the unique stationary distribution and the process approaches it as $\tau\to\infty$. The choice $B(z,z')=\max\{E(z),E(z')\}$ gives the Metropolis rates $M[z\to z']=\nu\min\{1,e^{-\beta[E(z')-E(z)]}\}$, and another common choice is the Glauber rates $\nu/(1+e^{\beta[E(z')-E(z)]})$, which also satisfy detailed balance. The graph and the barriers affect only the speed of relaxation; the stationary distribution, and hence all the bounds of this paper, depend only on the energy levels. The interaction time $\tau$ should be long compared with the relaxation time of the process (the inverse of the spectral gap of its generator); a finite $\tau$ is accounted for in Lemma~\ref{lem:finitetau}.

\begin{Definition}[Work capacity]\label{def:cap}
${\cal E}(m,s)$ is the class of all engines with segment length $m$, any number of stages $L$, and $s_1s_2\cdots s_L\le s$. Energies, rates (subject to detailed balance) and interaction times are arbitrary, and the engine may even be designed with knowledge of $x^N$. Define
\begin{equation}
C_{m,s}(x)=\liminf_{N\to\infty}\ \sup_{{\cal E}(m,s)}\frac{\langle W_N\rangle}{NkT},\qquad C(x)=\sup_{m,s}C_{m,s}(x),
\end{equation}
where $N$ ranges over multiples of $m$.
\end{Definition}

Two comments are in order. First, the number of stages is not restricted, and only the product $s_1\cdots s_L$ is bounded. The converse bound will hold for every $L$, and the engines of the achievability part will use many memoryless stages ($s_1=\cdots=s_L=1$), which keeps them in ${\cal E}(m,1)$. Second, the $\liminf_N$ is the dual of the $\limsup_N$ in the definition of finite--state compressibility \cite{ZL78}: in both cases, the worst lengths $N$ count. With this choice, $C(x)$ is determined by the $\limsup_N$ of empirical entropies, in the spirit of \cite{ZL78}.

\section{The Converse Bound}\label{sec:conv}

\subsection{One ratchet over many cycles}

Consider first a single cycle, with $\mu$ and $\mu'$ denoting the distributions of the Markov state at the beginning and at the end of the cycle. Since $E(\xi,\sigma)=kT\ln[1/\Pi(\xi,\sigma)]-kT\ln Z(\beta)$, the mean energy under any distribution $P$ of the Markov state is $kT[H(P)+D(P\|\Pi)]-kT\ln Z(\beta)$, and so the average work of the cycle is
\begin{align}
\frac{\langle W\rangle}{kT}&=\frac1{kT}\sum_{\xi,\sigma}\bigl[\mu'(\xi,\sigma)-\mu(\xi,\sigma)\bigr]E(\xi,\sigma)\nonumber\\
&=\bigl[H(\mu')-H(\mu)\bigr]-\bigl[D(\mu\|\Pi)-D(\mu'\|\Pi)\bigr].\label{eq:identity}
\end{align}
By detailed balance, $\Pi$ is a stationary distribution of the Markov process, and therefore $D(\mu'\|\Pi)\le D(\mu\|\Pi)$ (Lemma~\ref{lem:dp} in Appendix~\ref{app:markov}). Thus, the work is at most $kT$ times the entropy increase $H(\mu')-H(\mu)$, and the second bracket of \eqref{eq:identity} is the dissipation.

Now consider $K$ consecutive cycles, where the incoming segments $U^K=(U_1,\ldots,U_K)$ and the initial ratchet state $S_1$ have an arbitrary joint distribution. We claim that
\begin{equation}\label{eq:block}
\frac{\langle W\rangle}{kT}\le H(V^K,S_{K+1})-H(U^K,S_1).
\end{equation}
To see why, during cycle $j$ consider the extended Markov state $\Omega_t=(V^{j-1},\xi_t,\sigma_t,U_{j+1}^K)$, which also includes the segments already written and those not yet read. These components are frozen during cycle $j$, so $\{\Omega_t\}$ is a Markov process whose jumps change only $(\xi_t,\sigma_t)$, and the product of the uniform distribution on the frozen components with $\Pi$ is stationary for it. Since the frozen components carry no energy, \eqref{eq:identity} and the monotonicity of the divergence apply to $\{\Omega_t\}$ verbatim, and give $\langle W_j\rangle/kT\le H(\Omega_\tau)-H(\Omega_0)$ for the work $W_j$ of cycle $j$. (The uniform distribution here is only a reference measure; the true joint distribution of all components enters through the entropies.) Now, the end of cycle $j$ and the beginning of cycle $j+1$ consist of the same random variables, $(V^j,S_{j+1},U_{j+1}^K)$, so their entropies coincide, and summing over $j$, all intermediate terms cancel. This gives \eqref{eq:block}. Without the frozen components, the per--cycle bounds would not telescope, and the resulting bound would retain only per--segment entropies and a memory penalty $\ln s$ per cycle.

\subsection{From ordinary entropies to empirical entropies}

With a deterministic input, the input entropy in \eqref{eq:block} vanishes, and the bound is useless. The empirical entropy enters through the linearity of the work in the joint distribution of the input and the initial state, as follows.

Let $\ell$ be a multiple of $m$, so that each aligned $\ell$--block consists of $\ell/m$ whole segments, and consider a cascade. Let $N$ be a multiple of $\ell$, and $n=N/\ell$. For stage $r$ and block $i\in\{0,\ldots,n-1\}$, let
\begin{itemize}
\item $Y_r(i)$ be the output of stage $r$ on block $i$, with $Y_0(i)=\bx_i$;
\item $S_r(i)$ be the ratchet state of stage $r$ at the beginning of block $i$ (so that $S_r(n)$ is its final state);
\item $W_r(i)$ be the work of stage $r$ on block $i$.
\end{itemize}

\emph{(a) The work on a block depends only on how the block starts.} While processing block $i$, stage $r$ behaves in a way that depends only on its input block and its initial ratchet state. Hence there is a fixed function $\omega_r(y,\sigma)$, the expected work on a block that starts with input $y$ and state $\sigma$, such that
\begin{equation}\label{eq:omega}
\langle W_r(i)\rangle=\E\bigl\{\omega_r\bigl(Y_{r-1}(i),S_r(i)\bigr)\bigr\},
\end{equation}
and, similarly, the pair (output block, final state) is obtained from the pair (input block, initial state) by a fixed stochastic kernel.

\emph{(b) Averaging over blocks is the same as running one block from an averaged start.} Let $I$ be uniformly distributed over the blocks, independently of everything else, and let
\begin{equation}
\tilde Y_r=Y_r(I),\qquad \tilde S_r=S_r(I),\qquad \tilde S_r'=S_r(I+1).
\end{equation}
Averaging \eqref{eq:omega} over $i$ gives $\E\{\omega_r(\tilde Y_{r-1},\tilde S_r)\}$, which is the expected work of stage $r$ on a \emph{single} block whose input and initial ratchet state are drawn jointly from the distribution of $(\tilde Y_{r-1},\tilde S_r)$. The output and final ratchet state of that block are distributed as $(\tilde Y_r,\tilde S_r')$.

\emph{(c) The bound.} Applying \eqref{eq:block} to this single block, and then bounding $H(\tilde Y_r,\tilde S_r')\le H(\tilde Y_r)+\ln s_r$ and $H(\tilde Y_{r-1},\tilde S_r)\ge H(\tilde Y_{r-1})$,
\begin{equation}\label{eq:avg}
\frac1n\sum_{i=0}^{n-1}\langle W_r(i)\rangle\le kT\bigl[H(\tilde Y_r,\tilde S_r')-H(\tilde Y_{r-1},\tilde S_r)\bigr]\le kT\bigl[H(\tilde Y_r)-H(\tilde Y_{r-1})+\ln s_r\bigr].
\end{equation}

\subsection{The converse theorem}

Summing \eqref{eq:avg} over $r=1,\ldots,L$, the entropies of the intermediate tapes cancel (this is why the same index $I$ is used for all stages). Since $H(\tilde Y_0)=H(\bx_I)=\hat H_\ell(x^N)$ by \eqref{eq:emp}, $H(\tilde Y_L)\le\ell\ln|\X|$, and $\sum_r\ln s_r\le\ln s$, the total work is at most $nkT[\ell\ln|\X|-\hat H_\ell(x^N)+\ln s]$. Dividing by $N$, we have proved:

\begin{Theorem}[Converse]\label{thm:conv}
For every engine in ${\cal E}(m,s)$, every block length $\ell$ that is a multiple of $m$, and every $N$ that is a multiple of $\ell$,
\begin{equation}\label{eq:conv}
\frac{\langle W_N\rangle}{NkT}\le\ln|\X|-\frac{\hat H_\ell(x^N)}{\ell}+\frac{\ln s}{\ell}.
\end{equation}
Consequently, by \eqref{eq:H-LZ},
\begin{equation}\label{eq:convLZ}
\frac{\langle W_N\rangle}{NkT}\le\ln|\X|-\rho_{\rm LZ}(x^N)+\frac{\ln s}{\ell}+\delta(N,\ell).
\end{equation}
\end{Theorem}

Note that $\ell$ is a parameter of the analysis only; it does not affect the engine, and letting $\ell\to\infty$ removes the memory penalty $(\ln s)/\ell$.

\section{Universal Achievability}\label{sec:ach}

\subsection{The idea: quasi--static heating in space}

Consider a memoryless stage. Its Markov state is just the content $b\in\X^m$ of the segment it holds, and its design amounts to choosing one energy level $E(b)$ per possible content. If $\tau$ is large, the outgoing segment is a fresh sample from the canonical distribution $\Pi(b)\propto e^{-\beta E(b)}$, independent of the incoming one, and the work of the cycle is $E(\text{outgoing})-E(\text{incoming})$. Thus, the stage extracts work if the incoming segments sit at low energy levels. Writing $E(b)=kT\ln[1/Q(b)]$ for some probability distribution $Q$ on $\X^m$, this means that $Q$ should assign high probability to the segments that actually appear in $x$.

After such a stage, however, the tape is distributed according to $Q$, which may still have low entropy, and so most of the potential work is left unexploited. To extract it, we use $L$ stages with energy levels
\begin{equation}\label{eq:energies}
E_r(b)=\Bigl(1-\frac rL\Bigr)kT\ln\frac1{Q(b)},\qquad r=1,\ldots,L.
\end{equation}
Equivalently, stage $r$ has the energy levels $kT\ln[1/Q(b)]$ at temperature $T/(1-r/L)$, which grows gradually from $T$ to infinity along the cascade, where the distribution becomes uniform. Each stage then receives segments that are nearly in equilibrium with it, and so it dissipates very little. This is the spatial counterpart of a quasi--static process, which is normally defined in the time domain, and, as in the time domain, it is the only way to make the dissipation negligible. As we show next, the total work per segment is then approximately
\begin{equation}\label{eq:idea}
kT\Bigl[m\ln|\X|-\frac1J\sum_{i=0}^{J-1}\ln\frac1{Q(\bx_i)}\Bigr],
\end{equation}
where $J=N/m$ and $\bx_0,\ldots,\bx_{J-1}$ are the segments of $x^N$. The sum is the length, in nats, of the tape under a code with length function $\ln(1/Q)$. In other words, \emph{choosing the energy levels is choosing a lossless code}, and a universal code, fixed in advance, yields a universal engine.

\subsection{Analysis}

Let $Q$ be a distribution on $\X^m$ with $Q(b)>0$ for all $b$, and let
\begin{equation}\label{eq:path}
\Pi_\theta(b)=\frac{Q(b)^{1-\theta}}{\sum_{b'}Q(b')^{1-\theta}},\qquad 0\le\theta\le1,
\end{equation}
so that $\Pi_0=Q$ and $\Pi_1$ is uniform. The engine ${\cal U}[Q,L,\tau]$ consists of $L$ memoryless stages with energy levels \eqref{eq:energies}, detailed--balance rates under which the Markov jump process is irreducible (Appendix~\ref{app:markov}), and interaction time $\tau$; the canonical distribution of stage $r$ is $\Pi_{\theta_r}$, where $\theta_r=r/L$. We analyze the engine for $\tau=\infty$, i.e., when every stage relaxes completely; Lemma~\ref{lem:finitetau} in Appendix~\ref{app:markov} shows that a large finite $\tau$ costs at most an arbitrarily small $\gamma$ per symbol, uniformly in $x$ and $N$.

If the segments that enter stage $r$ have distribution (or empirical distribution) $P$, then, since each of them leaves with distribution $\Pi_{\theta_r}$, the average work of stage $r$ per segment is $\langle E_r\rangle_{\Pi_{\theta_r}}-\langle E_r\rangle_P$, i.e.,
\begin{equation}\label{eq:onestage}
kT\Bigl[H(\Pi_{\theta_r})-\sum_aP(a)\ln\frac1{\Pi_{\theta_r}(a)}\Bigr]=kT\bigl[H(\Pi_{\theta_r})-H(P)-D(P\|\Pi_{\theta_r})\bigr].
\end{equation}
Stage 1 receives the segments of $x^N$, with empirical distribution $\hat P_m$; this is the only place where $x$ enters. Stage $r\ge2$ receives i.i.d.\ segments with distribution $\Pi_{\theta_{r-1}}$. Summing \eqref{eq:onestage} over the stages, the entropies telescope to $H(\Pi_1)=m\ln|\X|$, and
\begin{equation}\label{eq:exact}
\frac{\langle W_N\rangle}{JkT}=m\ln|\X|-\sum_a\hat P_m(a)\ln\frac1{\Pi_{\theta_1}(a)}-\sum_{r=2}^LD(\Pi_{\theta_{r-1}}\|\Pi_{\theta_r}).
\end{equation}
The second term of \eqref{eq:exact} is the code length, up to a small overhead, and the third term is the total dissipation of stages $2,\ldots,L$. (By \eqref{eq:onestage}, the dissipation of the first stage is $D(\hat P_m\|\Pi_{\theta_1})$, an instance of the mismatch cost of \cite{KW17}.) Both losses are bounded by elementary estimates.

\emph{The first stage.} The second term of \eqref{eq:exact} is the average code length of the segments of $x^N$ under the landscape $\Pi_{\theta_1}$ of the first stage, which is $Q$ slightly flattened towards the uniform distribution. We compare it with the code length under $Q$ itself. By \eqref{eq:path}, for every $b$ and $\theta$,
\begin{equation}
\ln\frac1{\Pi_\theta(b)}=(1-\theta)\ln\frac1{Q(b)}+\ln\sum_{b'}Q(b')^{1-\theta}.
\end{equation}
The function $t\mapsto t^{1-\theta}$ is concave, so by Jensen's inequality,
\begin{equation}
\frac1{|\X|^m}\sum_{b'}Q(b')^{1-\theta}\le\Bigl(\frac1{|\X|^m}\sum_{b'}Q(b')\Bigr)^{1-\theta}=|\X|^{-m(1-\theta)},
\end{equation}
i.e., $\sum_{b'}Q(b')^{1-\theta}\le|\X|^{m\theta}$. Hence
\begin{equation}
\ln\frac1{\Pi_\theta(b)}\le(1-\theta)\ln\frac1{Q(b)}+\theta\,m\ln|\X|\le\ln\frac1{Q(b)}+\theta\,m\ln|\X|,
\end{equation}
where the second inequality holds because $\ln[1/Q(b)]\ge0$. Now take $\theta=\theta_1=1/L$ and average over $a\sim\hat P_m$. Since $\hat P_m$ is the empirical distribution of the segments $\bx_0,\ldots,\bx_{J-1}$, averaging a function of $a$ over $\hat P_m$ is the same as averaging it over the segments, and so
\begin{equation}\label{eq:first}
\sum_a\hat P_m(a)\ln\frac1{\Pi_{\theta_1}(a)}\le\frac1J\sum_{i=0}^{J-1}\ln\frac1{Q(\bx_i)}+\frac{m\ln|\X|}{L}.
\end{equation}
In words: flattening the landscape by $1/L$ costs at most $(m\ln|\X|)/L$ nats per segment.

\emph{The other stages.} $\{\Pi_\theta\}$ is an exponential family, and two members that are close in $\theta$ are close in divergence: $D(\Pi_\theta\|\Pi_{\theta'})\le\frac12(\theta'-\theta)^2\Lambda_m^2$ (Lemma~\ref{lem:expfam} in Appendix~\ref{app:lemmas}), where
\begin{equation}\label{eq:Lambda}
\Lambda_m=\max_b\bigl|\ln\bigl(|\X|^mQ(b)\bigr)\bigr|\le\max\Bigl\{m\ln|\X|,\ \max_b\ln\frac1{Q(b)}\Bigr\}.
\end{equation}
Here, $\Lambda_m$ bounds the absolute value of the statistic of the exponential family, which is $\ln[U(b)/Q(b)]$, where $U$ denotes the uniform distribution on $\X^m$. The inequality in \eqref{eq:Lambda} holds because
\begin{equation}
\ln\bigl(|\X|^mQ(b)\bigr)=m\ln|\X|-\ln\frac1{Q(b)}
\end{equation}
is the difference of two nonnegative numbers, and $|a-c|\le\max\{a,c\}$ for $a,c\ge0$. Since consecutive stages differ by $1/L$ in $\theta$,
\begin{equation}\label{eq:others}
\sum_{r=2}^LD(\Pi_{\theta_{r-1}}\|\Pi_{\theta_r})\le(L-1)\cdot\frac{\Lambda_m^2}{2L^2}\le\frac{\Lambda_m^2}{2L}.
\end{equation}
This is where the quasi--static character of the cascade enters: each step costs $O(1/L^2)$, and so $L$ steps cost $O(1/L)$. Substituting \eqref{eq:first} and \eqref{eq:others} into \eqref{eq:exact}, we have obtained:

\begin{Lemma}[Work as a code length]\label{lem:landscape}
For $\tau=\infty$, every $N$ that is a multiple of $m$, and every $x^N$, the engine ${\cal U}[Q,L,\infty]$ satisfies
\begin{equation}\label{eq:landscape}
\frac{\langle W_N\rangle}{JkT}\ge m\ln|\X|-\frac1J\sum_{i=0}^{J-1}\ln\frac1{Q(\bx_i)}-\frac{m\ln|\X|+\Lambda_m^2/2}{L}.
\end{equation}
\end{Lemma}

\subsection{A universal engine based on types}\label{sec:types}

Let $p$ and $q$ be positive integers and $m=qp$. Write each $b\in\X^m$ as a concatenation of $q$ blocks in $\X^p$, let $t(b)$ be the empirical distribution of these $q$ blocks (the \emph{type} of $b$), let ${\cal T}(b)$ be the set of all strings in $\X^m$ with the same type, and let ${\cal P}_q$ be the set of types. Define the two--part code (type, then index within the type class)
\begin{equation}\label{eq:Qp}
Q_p(b)=\frac1{|{\cal P}_q|\cdot|{\cal T}(b)|},\qquad b\in\X^m.
\end{equation}
Since $|{\cal P}_q|=\binom{q+A-1}{A-1}\le(q+1)^{A-1}$, with $A=|\X|^p$, and $|{\cal T}(b)|\le e^{qH(t(b))}$ \cite[Theorem~11.1.3]{CT06},
\begin{equation}\label{eq:Qp-bound}
\ln\frac1{Q_p(b)}\le qH(t(b))+(A-1)\ln(q+1).
\end{equation}
Since $p$ divides $m$, the aligned $p$--blocks of $x^N$ are exactly the $p$--blocks inside its segments, and so $\hat P_p$ is the average of the types $t(\bx_i)$. By the concavity of the entropy, $\frac1J\sum_iH(t(\bx_i))\le\hat H_p(x^N)$, and thus, by \eqref{eq:Qp-bound}, the code length in \eqref{eq:landscape} is at most $q\hat H_p(x^N)+(A-1)\ln(q+1)$. Also, by \eqref{eq:Lambda} and \eqref{eq:Qp-bound}, $\Lambda_m\le m\ln|\X|+(A-1)\ln(q+1)$. Substituting into Lemma~\ref{lem:landscape}, dividing by $N=mJ$, and accounting for a finite $\tau$ by Lemma~\ref{lem:finitetau}, we obtain the following result for the engine ${\cal U}[Q_p,L,\tau]$:

\begin{Theorem}[Universal achievability]\label{thm:ach}
For all positive integers $p,q,L$ and every $\gamma>0$, with $m=qp$, there is an engine in ${\cal E}(m,1)$, consisting of $L$ memoryless stages and not depending on $x$, such that for every $x$ and every $N$ that is a multiple of $m$,
\begin{equation}\label{eq:ach}
\frac{\langle W_N\rangle}{NkT}\ge\ln|\X|-\frac{\hat H_p(x^N)}{p}-\frac{(|\X|^p-1)\ln(q+1)}{m}-\frac1L\Bigl[\ln|\X|+\frac{\Lambda_m^2}{2m}\Bigr]-\gamma,
\end{equation}
where $\Lambda_m=m\ln|\X|+(|\X|^p-1)\ln(q+1)$.
\end{Theorem}

The energy levels of this engine have a simple structure: $E_r(b)=(1-\theta_r)kT\ln[|{\cal P}_q|\cdot|{\cal T}(b)|]$ depends on $b$ only through its type, i.e., through the numbers of occurrences of the various $p$--blocks in the segment. For example, for $|\X|=2$ and $p=1$, $|{\cal T}(b)|=\binom{m}{w(b)}$, where $w(b)$ is the number of 1's in $b$, so the first stage is a system of $m$ two--level cells whose energy depends only on the total number of 1's, $E_1(b)=(1-\frac1L)kT\ln[(m+1)\binom{m}{w(b)}]$. It is lowest for the all--zero and all--one segments and highest for balanced ones, so the stage extracts work from tapes with a biased symbol frequency, whichever symbol dominates. With single--flip moves in \eqref{eq:arrhenius}, each jump changes $w(b)$ by one.

\paragraph{Choice of the parameters.} The last three terms on the right--hand side of \eqref{eq:ach} are losses, and they are small only if $p$, $q$, $L$ and $\tau$ are chosen suitably, in this order. Given $p$, the third term, $[(|\X|^p-1)\ln(q+1)]/(pq)$, is small only if $q\gg|\X|^p\ln q$, i.e., the segment length $m=qp$ must grow exponentially with $p$. Given $p$ and $q$ with $q\ge|\X|^p$, we have $(|\X|^p-1)\ln(q+1)\le m\ln|\X|$, hence $\Lambda_m\le2m\ln|\X|$, so the fourth term is of order $m/L$, which is small if $L\gg m$. Finally, $\tau$ is chosen by Lemma~\ref{lem:finitetau} for the desired $\gamma$. Throughout the paper, we use the following \emph{standard choices}:
\begin{equation}\label{eq:standard}
L=m^3,\qquad \gamma=\frac1{m^2},\qquad q=|\X|^{2p},
\end{equation}
i.e., $\tau$ is chosen by Lemma~\ref{lem:finitetau} with $\gamma=1/m^2$, and, when the target is $\hat H_p(x^N)/p$, the segment length is $m=p|\X|^{2p}$.
With $q=|\X|^{2p}$, so that $m=p|\X|^{2p}$, we have $\ln(q+1)\le2p\ln|\X|+\ln2$, and so the third term of \eqref{eq:ach} is at most $(2\ln|\X|+\ln2)|\X|^{-p}$; since $q\ge|\X|^p$, $\Lambda_m\le2m\ln|\X|$, and so, with $L=m^3$, the fourth term is at most $m^{-3}[\ln|\X|+2m\ln^2|\X|]\le(\ln|\X|+2\ln^2|\X|)/m^2$. Since $m\ge|\X|^{2p}$, all the losses together are at most $c|\X|^{-p}$, where $c=3\ln|\X|+2\ln^2|\X|+\ln2+1$ depends only on $|\X|$:
\begin{equation}\label{eq:schedule}
\frac{\langle W_N\rangle}{NkT}\ge\ln|\X|-\frac{\hat H_p(x^N)}{p}-c|\X|^{-p}.
\end{equation}
Taking $\liminf_N$ and letting $p\to\infty$ along $p=k!$ (along which $\bar H_p(x)\to\bar H(x)$, Section~\ref{sec:prelim}), this single explicit sequence of engines extracts $kT[\ln|\X|-\bar H(x)]$ per symbol, for every $x$. The same standard choices are used for all the engines below (Proposition~\ref{prop:lzengine}, Theorems~\ref{thm:cap}, \ref{thm:univ}, \ref{thm:redund} and~\ref{thm:erase}), and in each case we verify that the losses vanish. The exponential growth of the segment length with $p$ is the thermodynamic counterpart of the fact that a universal code for the memoryless sources over the alphabet of $p$--blocks must be long compared with the number $|\X|^p-1$ of free parameters.

The third term on the right--hand side of \eqref{eq:ach} is the redundancy of the two--part code; in Section~\ref{sec:redund}, we show that the optimal redundancy of memoryless engines fixed in advance is about half of it. On incompressible tapes, the redundancy is paid in actual work consumed: the first energy landscape is not flat, and a tape that does not match it pays for that, just as a universal compressor slightly expands incompressible data. Finally, an engine designed for $x$ is obtained with $m=p$ and $Q=(1-\eta)\hat P_p+\eta\cdot$uniform: then the code length in \eqref{eq:landscape} is at most $\hat H_p(x^N)+\ln\frac1{1-\eta}$, with no redundancy, but the engine depends on $x$.

\subsection{A universal engine based on LZ78}\label{sec:lzland}

For $b\in\X^m$, let $b=w_1w_2\cdots w_c$ be its incremental parsing, where the last phrase $w_c$ is either new (complete) or equal to an earlier phrase (incomplete), and let
\begin{equation}\label{eq:lzlength}
{\rm LZ}(b)=\sum_{j=1}^{c}\ln j+\bigl(c-{\cal I}\{w_c\ \text{incomplete}\}\bigr)\ln|\X|
\end{equation}
be the length of the LZ78 code for $b$: phrase $j$ is described by a pointer to one of the $j$ strings $\varnothing,w_1,\ldots,w_{j-1}$ and, if complete, by its last symbol. These lengths satisfy the Kraft inequality, $Z_m=\sum_be^{-{\rm LZ}(b)}\le1$. Since ${\rm LZ}(b)$ is an idealized length, with no rounding to integers, this does not follow from the Kraft--McMillan inequality for an actual implementation of LZ78, but it follows from a probabilistic interpretation. Consider the random process that starts with the list $(\varnothing)$ and, at step $j=1,2,\ldots$, picks one of the $j$ entries of the current list and a symbol of $\X$, uniformly and independently, outputs their concatenation, and appends it to the list. Given the parsing $w_1\cdots w_c$ of $b$, let $F_b$ be the event that at each step $j\le c$ the process picks the first entry equal to $w_j$ with its last symbol deleted, together with that last symbol, except that at step $c$, if $w_c$ is incomplete, it picks the first entry equal to $w_c$ (and any symbol); by the definition of the incremental parsing, the required entries exist. Then $\Pr(F_b)=e^{-{\rm LZ}(b)}$, and, on $F_b$, the output of the process begins with $b$, so the events $F_b$, $b\in\X^m$, are disjoint, and hence $Z_m\le1$. Let $Q_{\rm LZ}(b)=e^{-{\rm LZ}(b)}/Z_m$. Then the energy levels \eqref{eq:energies} of ${\cal U}[Q_{\rm LZ},L,\tau]$ are $(1-r/L)kT\,{\rm LZ}(b)$, up to additive constants: \emph{the energy levels are scaled LZ78 code lengths}. Since $\ln[1/Q_{\rm LZ}(b)]\le{\rm LZ}(b)\le c\ln c+c\ln|\X|$, the code length in \eqref{eq:landscape} is essentially the block LZ complexity of $x^N$:

\begin{Proposition}[LZ engines]\label{prop:lzengine}
For every $m$ there is an engine ${\cal U}^{\rm LZ}_m\in{\cal E}(m,1)$, not depending on $x$, whose stage energies are scaled LZ78 code lengths of the segment contents, such that for every $x$ and every $N$ that is a multiple of $m$,
\begin{equation}\label{eq:lzengine}
\frac{\langle W_N({\cal U}^{\rm LZ}_m)\rangle}{NkT}\ge\ln|\X|-\frac mN\sum_{i=0}^{N/m-1}\rho_{\rm LZ}(\bx_i)-\varepsilon_m,
\end{equation}
where $\bx_i$ are the segments of $x^N$, and $\varepsilon_m\to0$ as $m\to\infty$, independently of $x$ and $N$.
\end{Proposition}

\begin{proof}
Let $c_m$ be the maximal number of phrases in the incremental parsing of a string of length $m$; then
\begin{equation}
c_m\le\frac{m\ln|\X|}{(1-\epsilon'_m)\ln m},
\end{equation}
with $\epsilon'_m\to0$ \cite{ZL78}, and so $\zeta_m=(c_m\ln|\X|)/m\to0$. Since $c$ in \eqref{eq:lzlength} is the number of phrases $c(b)$ of Section~\ref{sec:prelim} (both count the last phrase even if it is incomplete), $\frac1m\ln[1/Q_{\rm LZ}(b)]\le\frac1m{\rm LZ}(b)\le\rho_{\rm LZ}(b)+\zeta_m$ for every segment $b$. Also, ${\rm LZ}(b)\le m\ln(m|\X|)$, so that $\Lambda_m\le m\ln(m|\X|)$ by \eqref{eq:Lambda}. Take ${\cal U}^{\rm LZ}_m={\cal U}[Q_{\rm LZ},m^3,\tau_m]$, with $\tau_m$ chosen by Lemma~\ref{lem:finitetau} with $\gamma=1/m^2$, in accordance with the standard choices \eqref{eq:standard} (there is no $p$ here). Lemma~\ref{lem:landscape} then gives \eqref{eq:lzengine} with $\varepsilon_m=\zeta_m+[\ln|\X|+\frac12m\ln^2(m|\X|)]/m^3+1/m^2$, which tends to zero since $\zeta_m\to0$ and $m\ln^2(m|\X|)/m^3\to0$.
\end{proof}

\section{The Work Capacity}\label{sec:cap}

Combining Theorems~\ref{thm:conv} and~\ref{thm:ach}, we obtain an exact characterization of the work capacity.

\begin{Theorem}[Work capacity]\label{thm:cap}
For every infinite sequence $x$,
\begin{equation}\label{eq:cap}
C(x)=\ln|\X|-\bar H(x).
\end{equation}
\end{Theorem}

\begin{proof}
\emph{Upper bound.} Fix $m$ and $s$, and apply \eqref{eq:conv} with $\ell=jm$. The right--hand side does not depend on the engine, so it also bounds the supremum over ${\cal E}(m,s)$. Since a $\liminf_N$ over the multiples of $m$ is at most the $\liminf_N$ over the multiples of $jm$, taking the latter gives
\begin{equation}
C_{m,s}(x)\le\ln|\X|-\bar H_{jm}(x)+\frac{\ln s}{jm}.
\end{equation}
Now let $j\to\infty$: the class ${\cal E}(m,s)$ is fixed, and only the analysis parameter $\ell=jm$ grows. Since $\bar H_{jm}(x)\ge\bar H(x)$ for every $j$, we get $C_{m,s}(x)\le\ln|\X|-\bar H(x)$ for all $m,s$.

\emph{Lower bound.} Fix $p$, and take the engine of Theorem~\ref{thm:ach} with the standard choices \eqref{eq:standard}, so that $m=p|\X|^{2p}$. It belongs to ${\cal E}(m,1)$, so, by \eqref{eq:schedule}, taking $\liminf_N$,
\begin{equation}
C(x)\ge C_{m,1}(x)\ge\ln|\X|-\bar H_p(x)-c|\X|^{-p}.
\end{equation}
Letting $p\to\infty$ along $p=k!$, along which $\bar H_p(x)\to\bar H(x)$, gives $C(x)\ge\ln|\X|-\bar H(x)$.
\end{proof}

The engines that attain the capacity need not be designed for $x$. A single engine of Theorem~\ref{thm:ach} serves one block length $p$; to serve all of them simultaneously, we mix the type--class codes over all divisors of $m$.

\begin{Theorem}[Universal engines]\label{thm:univ}
There is a sequence of engines ${\cal U}_m\in{\cal E}(m,1)$, $m=1,2,\ldots$, none of which depends on $x$, such that for every infinite sequence $x$,
\begin{equation}\label{eq:univ}
\lim_{j\to\infty}\ \liminf_{N\to\infty}\frac{\langle W_N({\cal U}_{j!})\rangle}{NkT}=C(x),
\end{equation}
where $N$ ranges over multiples of $j!$.
\end{Theorem}

\begin{proof}
\emph{The engines.} Let $Q^*_m$ be the uniform mixture of the type--class codes $Q_p$ of \eqref{eq:Qp} over all divisors $p$ of $m$. Since $m$ has at most $m$ divisors,
\begin{equation}
\ln\frac1{Q^*_m(b)}\le\ln\frac1{Q_p(b)}+\ln m\qquad\text{for every divisor $p$ of $m$}.
\end{equation}
In particular (take $p=1$ in \eqref{eq:Qp-bound}), $\Lambda_m\le\Lambda^*_m=m\ln|\X|+(|\X|-1)\ln(m+1)+\ln m$. Let ${\cal U}_m={\cal U}[Q^*_m,m^3,\tau_m]$, with $\tau_m$ chosen by Lemma~\ref{lem:finitetau} with $\gamma=1/m^2$, in accordance with the standard choices \eqref{eq:standard}.

\emph{A bound for every divisor.} As in the derivation of Theorem~\ref{thm:ach}, for every divisor $p$ of $m$, every $x$, and every $N$ that is a multiple of $m$,
\begin{align}
\frac{\langle W_N({\cal U}_m)\rangle}{NkT}\ge{}&\ln|\X|-\frac{\hat H_p(x^N)}{p}-\frac{(|\X|^p-1)\ln(m/p+1)+\ln m}{m}\nonumber\\
&-\frac1{m^3}\Bigl[\ln|\X|+\frac{(\Lambda^*_m)^2}{2m}\Bigr]-\frac1{m^2}.\label{eq:univbound}
\end{align}

\emph{Lower bound.} Fix $p$. For every $j\ge p$, $p$ divides $m=j!$, so \eqref{eq:univbound} applies. Take $\liminf_N$ and then let $j\to\infty$. For fixed $p$, the third term of \eqref{eq:univbound} is $O\bigl((|\X|^p\ln m)/m\bigr)$, and, since $\Lambda^*_m=O(m)$, the fourth and fifth terms are $O(1/m^2)$; so the last three terms vanish. Hence the left--hand side of \eqref{eq:univ} is at least $\ln|\X|-\bar H_p(x)$, for every $p$, and therefore at least $\ln|\X|-\bar H(x)=C(x)$.

\emph{Upper bound.} ${\cal U}_m\in{\cal E}(m,1)$, so $\liminf_N\langle W_N({\cal U}_m)\rangle/(NkT)\le C_{m,1}(x)\le C(x)$.
\end{proof}

The subsequence $m=j!$ only ensures that every $p$ eventually divides $m$. Each engine ${\cal U}_m$ is finite, and $C(x)$ is attained for every $x$ only in the limit $m\to\infty$, which parallels the fact that no single finite--state encoder attains the finite--state compressibility of every sequence, whereas the LZ algorithm does.

Finally, the capacity can be expressed in terms of the LZ complexity. The content of the next proposition is essentially known: by \cite[eq.~(38)]{p246}, $\lim_m\rho_m(x)=\bar H_{\rm csw}[x]$, and the equality between the entropy rates of aligned and non--aligned blocks is known from the literature on finite--state dimension \cite{BHV05,KS21}, where the normalized block entropies with $\liminf_N$ and $\limsup_N$ are the finite--state dimension and the strong finite--state dimension, respectively (up to the factor $\ln|\X|$), and the non--aligned blocks are taken along a non--cyclic sliding window. We include a short self--contained proof in the present notation, in which one of the two inequalities is obtained thermodynamically, from the LZ engine of Proposition~\ref{prop:lzengine} and the converse.

\begin{Proposition}[LZ characterization]\label{prop:lz}
For every infinite sequence $x$,
\begin{equation}
\bar H(x)=\bar H_{\rm csw}[x]=\lim_{m\to\infty}\rho_m(x),
\end{equation}
and so $C(x)=\ln|\X|-\lim_{m\to\infty}\rho_m(x)$.
\end{Proposition}

\begin{proof}
By \eqref{eq:p246}, it suffices to show that $\lim_m\rho_m(x)=\bar H(x)$; the expression for $C(x)$ then follows from Theorem~\ref{thm:cap}. We prove the two inequalities separately; the first one is thermodynamic.

\emph{$\bar H(x)\le\liminf_m\rho_m(x)$.} Fix $m$. Taking $\liminf_N$ in \eqref{eq:lzengine}, the LZ engine ${\cal U}^{\rm LZ}_m$ satisfies $\liminf_N\langle W_N\rangle/(NkT)\ge\ln|\X|-\rho_m(x)-\varepsilon_m$. On the other hand, ${\cal U}^{\rm LZ}_m\in{\cal E}(m,1)$, and so, by the upper bound in the proof of Theorem~\ref{thm:cap}, the same $\liminf_N$ is at most $C_{m,1}(x)\le\ln|\X|-\bar H(x)$. Hence $\bar H(x)\le\rho_m(x)+\varepsilon_m$ for every $m$, and since $\varepsilon_m\to0$, $\bar H(x)\le\liminf_m\rho_m(x)$. (Proposition~\ref{prop:lzengine} and the upper bound of Theorem~\ref{thm:cap} do not rely on the present proposition, so there is no circularity.)

\emph{$\lim_m\rho_m(x)\le\bar H(x)$.} Fix $\ell$, let $m$ be a multiple of $\ell$, let $N$ be a multiple of $m$, and let $\boldsymbol{b}_0,\ldots,\boldsymbol{b}_{N/m-1}$ be the aligned $m$--blocks of $x^N$. Applying \eqref{eq:H-LZ} to each $\boldsymbol{b}_i$, which is a string of length $m$, gives $\rho_{\rm LZ}(\boldsymbol{b}_i)\le\hat H_\ell(\boldsymbol{b}_i)/\ell+\delta(m,\ell)$. Since $\ell$ divides $m$, the aligned $\ell$--blocks of $x^N$ are exactly the aligned $\ell$--blocks inside the $\boldsymbol{b}_i$, so the empirical distribution of $\ell$--blocks of $x^N$ is the average of those of the $\boldsymbol{b}_i$, and by the concavity of the entropy,
\begin{equation}
\frac mN\sum_{i=0}^{N/m-1}\rho_{\rm LZ}(\boldsymbol{b}_i)\le\frac mN\sum_{i=0}^{N/m-1}\frac{\hat H_\ell(\boldsymbol{b}_i)}{\ell}+\delta(m,\ell)\le\frac{\hat H_\ell(x^N)}{\ell}+\delta(m,\ell).
\end{equation}
Taking $\limsup_N$ over the multiples of $m$, $\rho_m(x)\le \bar H_\ell(x)+\delta(m,\ell)$. As $m\to\infty$ along the multiples of $\ell$, $\delta(m,\ell)\to(\ln2)/\ell$. Since the limit $\lim_m\rho_m(x)$ exists \cite[eq.~(38)]{p246}, it coincides with the limit along the multiples of $\ell$, and so $\lim_m\rho_m(x)\le \bar H_\ell(x)+(\ln2)/\ell$ for every $\ell$. Applying this with $k\ell$ in place of $\ell$, using $\bar H_{k\ell}(x)\le \bar H_\ell(x)$, and letting $k\to\infty$, gives $\lim_m\rho_m(x)\le \bar H_\ell(x)$ for every $\ell$, that is, $\lim_m\rho_m(x)\le\bar H(x)$.
\end{proof}

Without using the existence of $\lim_m\rho_m(x)$ from \cite{p246}, the same argument shows that $\liminf_m\rho_m(x)=\bar H(x)$; thus \cite{p246} is needed only for the existence of the limit.

\begin{Corollary}[LZ engines attain the capacity]\label{cor:lzengine}
The engines ${\cal U}^{\rm LZ}_m$ of Proposition~\ref{prop:lzengine} satisfy, for every infinite sequence $x$,
\begin{equation}
\lim_{m\to\infty}\ \liminf_{N\to\infty}\frac{\langle W_N({\cal U}^{\rm LZ}_m)\rangle}{NkT}=C(x).
\end{equation}
\end{Corollary}

\begin{proof}
Taking $\liminf_N$ in \eqref{eq:lzengine} gives at least $\ln|\X|-\rho_m(x)-\varepsilon_m$, which tends to $\ln|\X|-\bar H(x)=C(x)$ by Proposition~\ref{prop:lz}. Conversely, ${\cal U}^{\rm LZ}_m\in{\cal E}(m,1)$, so the $\liminf_N$ is at most $C(x)$.
\end{proof}

\section{A Kraft Inequality for Work and the Optimal Redundancy}\label{sec:redund}

\subsection{The question}\label{sec:question}

An engine is \emph{fixed in advance} if its energy levels, rates and interaction times do not depend on the input. The engines of Theorems~\ref{thm:ach} and~\ref{thm:univ} are fixed in advance, but they pay a price for it. Fix integers $p,q\ge1$, let $m=qp$, and, for an engine and an input $x^N$ ($N$ a multiple of $m$), define the \emph{shortfall}
\begin{equation}\label{eq:shortfall}
\Delta(x^N)=\ln|\X|-\frac{\hat H_p(x^N)}p-\frac{\langle W_N(x^N)\rangle}{NkT},
\end{equation}
i.e., the gap, per symbol and in units of $kT$, between the target $\ln|\X|-\hat H_p(x^N)/p$ and the work actually extracted. By Theorem~\ref{thm:ach}, the engine ${\cal U}[Q_p,L,\tau]$ has $\Delta(x^N)\lesssim[(|\X|^p-1)\ln(q+1)]/m$ for every $x^N$. Can another engine fixed in advance do better? In universal source coding, the analogous question has a classical answer: for every lossless code fixed in advance, the worst--case excess of the code length over the empirical entropy is at least the logarithm of the Shtarkov sum \cite{Shtarkov87}. In this section, we show that the same answer holds for memoryless engines. The bridge between the two problems is the following observation: \emph{in an engine fixed in advance, $\ln|\X|$ per symbol minus the work behaves like a code length}, i.e., it satisfies the Kraft inequality. This is the converse of Lemma~\ref{lem:landscape}, which says that an engine built from a code extracts $\ln|\X|$ per symbol minus the code length.

\subsection{A Kraft inequality for work}

We begin with a single cycle of a single ratchet. Let ${\cal Z}$ be its set of Markov states, $E(z)$ its energy levels, and $P(z,z')$ the probability that the Markov state at the end of the cycle is $z'$, given that it is $z$ at the beginning. By the definition of the work in Section~\ref{sec:model}, the work of the cycle is $W=E(z')-E(z)$, whatever the path between $z$ and $z'$. Denote by $\langle\cdot\rangle_z$ the expectation given the initial Markov state $z$.

\begin{Lemma}\label{lem:ift}
For one cycle of a ratchet, $\sum_{z\in{\cal Z}}\langle e^{\beta W}\rangle_z=|{\cal Z}|$.
\end{Lemma}

\begin{proof}
By detailed balance, $e^{-\beta E(z)}P(z,z')=e^{-\beta E(z')}P(z',z)$ for all $z,z'$ (Lemma~\ref{lem:revP} in Appendix~\ref{app:markov}), i.e., $P(z,z')e^{\beta[E(z')-E(z)]}=P(z',z)$. Hence
\begin{equation}
\sum_z\langle e^{\beta W}\rangle_z=\sum_z\sum_{z'}P(z,z')e^{\beta[E(z')-E(z)]}=\sum_{z'}\sum_zP(z',z)=\sum_{z'}1=|{\cal Z}|.
\end{equation}
\end{proof}

In words: if the initial Markov state is uniformly distributed, then $\langle e^{\beta W}\rangle=1$. Since the heat delivered to the bath is $-W$, the quantity $-\beta W$ is the entropy production in the medium, and the identity is an instance of the integral fluctuation theorems of stochastic thermodynamics \cite{Seifert05}, with the uniform distribution serving both as the initial distribution and as the final reference distribution; it is in the spirit of the Jarzynski equality \cite{Jarzynski97}. Note that it holds for every interaction time, not only for $\tau=\infty$.

Now consider a memoryless engine, i.e., an engine in ${\cal E}(m,1)$: each stage $r$ has a single ratchet state, so its Markov state is just the segment content $b\in\X^m$, and its energy levels are $E_r(b)$. Then the stages act on each segment separately, and the expected work on $x^N$ is $\langle W_N(x^N)\rangle=\sum_i\langle W_m(\bx_i)\rangle$, where $\langle W_m(b)\rangle$ denotes the expected work of the whole cascade on a single segment with content $b$.

\begin{Lemma}[Kraft inequality for work]\label{lem:kraftwork}
For every memoryless engine, the numbers
\begin{equation}\label{eq:worklength}
{\cal L}(b)=m\ln|\X|-\frac{\langle W_m(b)\rangle}{kT},\qquad b\in\X^m,
\end{equation}
satisfy $\sum_{b\in\X^m}e^{-{\cal L}(b)}\le1$.
\end{Lemma}

\begin{proof}
Let $b_0=b$ be the input segment and $b_r$ the segment after stage $r$, $r=1,\ldots,L$. The work of stage $r$ is $E_r(b_r)-E_r(b_{r-1})$, and given $b_{r-1}$, the segment $b_r$ is drawn from $P_r(b_{r-1},\cdot)$, the transition matrix of stage $r$ over one cycle, independently of $b_0,\ldots,b_{r-2}$. Hence, by the same identity as in the proof of Lemma~\ref{lem:ift}, applied to each stage,
\begin{equation}
\bigl\langle e^{\beta W_m}\bigr\rangle_{b_0}=\sum_{b_1,\ldots,b_L}\prod_{r=1}^LP_r(b_{r-1},b_r)e^{\beta[E_r(b_r)-E_r(b_{r-1})]}=\sum_{b_1,\ldots,b_L}\prod_{r=1}^LP_r(b_r,b_{r-1}).
\end{equation}
Summing over $b_0$, then over $b_1$, and so on up to $b_{L-1}$, each sum $\sum_{b_{r-1}}P_r(b_r,b_{r-1})$ equals $1$, and the remaining sum over $b_L$ gives $|\X|^m$. Thus $\sum_b\langle e^{\beta W_m}\rangle_b=|\X|^m$. Since $e^{\beta\langle W_m\rangle_b}\le\langle e^{\beta W_m}\rangle_b$ by Jensen's inequality, $\sum_be^{\beta\langle W_m(b)\rangle}\le|\X|^m$, which is the assertion.
\end{proof}

Jensen's inequality discards information: the identity $\sum_b\langle e^{\beta W_m}\rangle_b=|\X|^m$, established in the proof, constrains the fluctuations of the work, not only its mean. For example, for every function ${\cal L}'$ on $\X^m$ that satisfies the Kraft inequality and every $a>0$, Markov's inequality gives
\begin{equation}
\sum_{b\in\X^m}e^{-{\cal L}'(b)}\Pr\Bigl\{W_m(b)\ge kT\bigl[m\ln|\X|-{\cal L}'(b)+a\bigr]\Bigr\}\le e^{-a}\sum_{b\in\X^m}|\X|^{-m}\bigl\langle e^{\beta W_m}\bigr\rangle_b=e^{-a},
\end{equation}
where $W_m(b)$ is the (random) work on a segment with content $b$: on segments weighted by any code, the work exceeds $kT$ times $m\ln|\X|$ minus the code length by $a$ nats only with probability of order $e^{-a}$.

\subsection{The optimal redundancy}

We now use Lemma~\ref{lem:kraftwork} to translate the question of Section~\ref{sec:question} into a question about codes. Let the input be periodic: $x^N$ consists of $N/m$ repetitions of a single segment $b$. Then two things happen. First, every segment of $x^N$ is $b$, so the expected work is $(N/m)\langle W_m(b)\rangle$, and, by \eqref{eq:worklength}, $\ln|\X|-\langle W_N(x^N)\rangle/(NkT)={\cal L}(b)/m$. Second, the aligned $p$--blocks of $x^N$ are those of $b$, repeated $N/m$ times, so $\hat H_p(x^N)=H(t(b))$, the entropy of the type of $b$ (Section~\ref{sec:types}). Substituting both in \eqref{eq:shortfall},
\begin{equation}\label{eq:periodic}
\Delta(x^N)=\frac1m\bigl[{\cal L}(b)-qH(t(b))\bigr].
\end{equation}
The bracket has a familiar meaning. View $b$ as a sequence of $q$ symbols (its $p$--blocks) from an alphabet of size $A=|\X|^p$. Then $qH(t(b))$ is the ideal code length of $b$ under the memoryless source that fits $b$ best, i.e., $e^{-qH(t(b))}$ is the largest probability that a memoryless source on $\X^p$ can assign to $b$. And ${\cal L}(\cdot)$ satisfies the Kraft inequality (Lemma~\ref{lem:kraftwork}), so it is the length function of a lossless code (real--valued, which is all that the converses below use). Thus, the bracket in \eqref{eq:periodic} is the \emph{regret} of the code ${\cal L}$ on $b$, relative to the class of memoryless sources, and every lower bound on the regret of codes is also a lower bound on the shortfall of memoryless engines.

The classical such bound is Shtarkov's \cite{Shtarkov87}: for every function ${\cal L}$ on $\X^m$ that satisfies the Kraft inequality,
\begin{equation}\label{eq:shtarkovbound}
\max_{b\in\X^m}\bigl[{\cal L}(b)-qH(t(b))\bigr]\ge\ln S_q(A),\qquad S_q(A)=\sum_{b\in\X^m}e^{-qH(t(b))},
\end{equation}
where $S_q(A)$ is the \emph{Shtarkov sum}. (Indeed, if $R$ denotes the left--hand side, then $e^{-qH(t(b))}\le e^Re^{-{\cal L}(b)}$ for every $b$, and summing over $b$ gives $S_q(A)\le e^R$.) The bound is tight: equality holds for the normalized maximum--likelihood code, ${\cal L}(b)=qH(t(b))+\ln S_q(A)$. As $q\to\infty$ \cite{XB00},
\begin{equation}\label{eq:shtarkovasym}
\ln S_q(A)=\frac{A-1}2\ln\frac q{2\pi}+\ln\frac{\Gamma(1/2)^A}{\Gamma(A/2)}+o(1).
\end{equation}

\begin{Theorem}[Optimal redundancy of memoryless universal engines]\label{thm:redund}
Let $p$ and $q$ be positive integers, $m=qp$ and $A=|\X|^p$.
\begin{enumerate}[label=(\alph*),leftmargin=*]
\item For every memoryless engine fixed in advance and every $N$ that is a multiple of $m$, there is an input $x^N$ with shortfall $\Delta(x^N)\ge[\ln S_q(A)]/m$.
\item For every positive integer $L$ and every $\gamma>0$, there is a memoryless engine with $L$ stages, fixed in advance, such that $\Delta(x^N)\le[\ln S_q(A)]/m+(1/L)[\ln|\X|+\Lambda_m^2/(2m)]+\gamma$ for every $x$ and every $N$ that is a multiple of $m$, where $\Lambda_m=m\ln|\X|+\ln S_q(A)$.
\end{enumerate}
\end{Theorem}

\begin{proof}
(a) By \eqref{eq:shtarkovbound}, there is a segment $b$ with ${\cal L}(b)-qH(t(b))\ge\ln S_q(A)$, and, by \eqref{eq:periodic}, the input $x^N$ that consists of $N/m$ repetitions of $b$ has $\Delta(x^N)\ge[\ln S_q(A)]/m$.

(b) Let $Q_{\rm NML}$ be the normalized maximum--likelihood distribution, which corresponds to the code that attains equality in \eqref{eq:shtarkovbound},
\begin{equation}
Q_{\rm NML}(b)=\frac{e^{-qH(t(b))}}{S_q(A)},\qquad b\in\X^m,
\end{equation}
so that $\ln[1/Q_{\rm NML}(b)]=qH(t(b))+\ln S_q(A)$ exactly, and take the engine ${\cal U}[Q_{\rm NML},L,\tau]$ of Section~\ref{sec:ach}. The rest is as in the proof of Theorem~\ref{thm:ach}: by the concavity of the entropy, the code length in \eqref{eq:landscape} is at most $q\hat H_p(x^N)+\ln S_q(A)$; by \eqref{eq:Lambda}, $\Lambda_m\le m\ln|\X|+\ln S_q(A)$, since $qH(t(b))\le m\ln|\X|$; and Lemmas~\ref{lem:landscape} and~\ref{lem:finitetau} give the bound.
\end{proof}

In Theorem~\ref{thm:redund}(b), the first term on the right--hand side is the optimal redundancy, and the other two are losses. With the standard choices \eqref{eq:standard}, they are negligible. Indeed, $S_q(A)=\sum_t|{\cal T}_t|e^{-qH(t)}\le|{\cal P}_q|\le(q+1)^{A-1}$, where the sum is over the types $t$ and ${\cal T}_t$ is the corresponding type class, and $(A-1)\ln(q+1)\le m\ln|\X|$ for $q\ge|\X|^p$, which holds for $q=|\X|^{2p}$; hence $\Lambda_m\le2m\ln|\X|$, and with $L=m^3$ and $\gamma=1/m^2$, the losses are at most $(1+2m\ln|\X|)\ln|\X|/m^3+1/m^2=O(1/m^2)$, which is negligible compared with $[\ln S_q(A)]/m$.

Part (a) of Theorem~\ref{thm:redund} is a minimax statement: for every engine, \emph{some} input has a large shortfall. To prove it, it suffices to exhibit one such input, and the periodic input built from a worst segment serves as a witness; the restriction to periodic inputs in the proof therefore entails no loss of generality. Part (b), on the other hand, holds for \emph{every} input. Together, they determine the worst case of the shortfall over all inputs $x^N$: up to the losses, which are $O(1/m^2)$ under the standard choices, the best worst--case shortfall of memoryless engines fixed in advance is exactly $[\ln S_q(A)]/m$ per symbol. By \eqref{eq:shtarkovasym}, this is about $[(|\X|^p-1)\ln q]/(2m)$, half the redundancy of the two--part code of Theorem~\ref{thm:ach}. The optimal engine has a transparent structure: its energy levels, $E_r(b)=(1-\theta_r)kT[qH(t(b))+\ln S_q(A)]$, are proportional to the empirical entropy of the segment, up to an additive constant.

In fact, the shortfall $[(A-1)\ln q]/(2m)$ cannot be avoided on most periodic inputs. This follows by applying, through \eqref{eq:periodic}, the individual--sequence lower bound of Weinberger, Merhav and Feder \cite{WMF94}, which is the deterministic counterpart of Rissanen's lower bound for probabilistic sources, and which, like \eqref{eq:shtarkovbound}, holds for every probability assignment, hence for every length function that satisfies the Kraft inequality.

\begin{Corollary}[Shortfall on most periodic inputs]\label{cor:most}
Fix $p$ and $\epsilon>0$, let $A=|\X|^p$, and, for every $q$, let $m=qp$ and consider any memoryless engine fixed in advance. Let ${\cal B}_q$ be the set of segments $b\in\X^m$ for which the periodic input that consists of $N/m$ repetitions of $b$ has shortfall
\begin{equation}
\Delta(x^N)<\Bigl(\frac{A-1}2-\epsilon\Bigr)\frac{\ln q}m,
\end{equation}
and, viewing segments as sequences of $q$ symbols from an alphabet of size $A$, call a type $t$ \emph{bad} if more than a fraction $q^{-\epsilon/3}$ of the segments of type $t$ belong to ${\cal B}_q$. Then the fraction of bad types among all types tends to zero as $q\to\infty$.
\end{Corollary}

\begin{proof}
By \eqref{eq:periodic}, $b\in{\cal B}_q$ if and only if ${\cal L}(b)<qH(t(b))+[(A-1)/2-\epsilon]\ln q$. Let $P(b)=e^{-{\cal L}(b)}$, which satisfies $\sum_bP(b)\le1$ by Lemma~\ref{lem:kraftwork}. Apply \cite[Theorem~1]{WMF94} to the probability assignment $P$ on sequences of length $q$ over the alphabet $\X^p$ of size $A$, with the finite--state machine that has a single state ($k=1$), for which the empirical conditional entropy is $H(t(b))$ and the finite--state types are the types $t$. In natural logarithms, the exceptional set of \cite[Theorem~1]{WMF94}, defined by $-\ln P(b)<qH(t(b))+[k(A-1)/2-\epsilon]\ln q$, is ${\cal B}_q$, and \cite[Theorem~1]{WMF94} asserts that the fraction of types in which ${\cal B}_q$ occupies more than a fraction $q^{-\epsilon/3}$ vanishes as $q\to\infty$. The engine, and hence $P$, may change with $q$. This is allowed, because the proof of \cite[Theorem~1]{WMF94} bounds the fraction of bad types by the sum of two quantities that do not depend on the probability assignment: the bound of \cite[Corollary~1]{WMF94}, which applies to every probability assignment, and the vanishing fraction of atypical types of \cite[Lemma~4]{WMF94}. Hence the convergence holds for every choice of engines, one for each $q$.
\end{proof}

Unlike Theorem~\ref{thm:redund}(a), where periodic inputs serve only as witnesses, Corollary~\ref{cor:most} is a statement about the periodic inputs themselves, and it does not assert that the shortfall is large on most inputs in general. The restriction is natural: the work of a memoryless engine is additive over the segments, whereas the target $\hat H_p(x^N)$ is a concave function of the empirical distribution of the whole input, and periodic inputs are exactly those on which the two coincide segment by segment, so that the engine problem becomes a coding problem on a single segment. Extending Corollary~\ref{cor:most} to general inputs is not immediate, for two reasons. First, for an input with segments $b_1,\ldots,b_J$,
\begin{equation}
\Delta(x^N)=\frac1J\sum_{i=1}^J\frac{{\cal L}(b_i)-qH(t(b_i))}m-\frac1p\Bigl[\hat H_p(x^N)-\frac1J\sum_{i=1}^JH(t(b_i))\Bigr],
\end{equation}
and the bracket, which is nonnegative by the concavity of the entropy and vanishes for periodic inputs, reduces the shortfall; for typical non--periodic inputs, it may be of the same order as the regret itself. Second, even when all the segments have the same type, so that the bracket vanishes, the shortfall is an average of the regrets of the segments, and the small fraction of segments with small (possibly negative) regret that \cite{WMF94} permits may dominate this average. The same reduction applies to other reference classes, with the target and the regret defined segment by segment; for example, for finite--state reference models (with $k>1$ states and a strongly connected graph), \cite[Theorem~1]{WMF94} gives the corresponding lower bound, with $(A-1)/2$ replaced by $k(A-1)/2$. We do not pursue this here.

\paragraph{Engines with memory.} The argument of Lemma~\ref{lem:kraftwork} extends to engines with memory, if it is applied to the global configuration that consists of the contents of all segments and the ratchet states of all stages, rather than to a single segment. Since the ratchet states need not start uniformly distributed, the bound becomes $s$ instead of $1$: for every engine in ${\cal E}(m,s)$, every initial distribution of its ratchet states, and every $N$ that is a multiple of $m$,
\begin{equation}\label{eq:kraftwork}
\sum_{x^N\in\X^N}\exp\Bigl\{\frac{\langle W_N(x^N)\rangle}{kT}-N\ln|\X|\Bigr\}\le s
\end{equation}
(Lemma~\ref{lem:kraftmem} in Appendix~\ref{app:kraftmem}). Now the Kraft--type constraint is on the whole input $x^N$, not on each segment separately, and the argument of Theorem~\ref{thm:redund}(a) yields only a shortfall of at least $[\ln S_{N/p}(A)-\ln s]/N$, which vanishes as $N\to\infty$. Whether memory can reduce the redundancy below $[\ln S_q(A)]/m$ is open.

\section{Incompressible Sequences and Erasure}\label{sec:erase}

The two bounds of Sections~\ref{sec:conv} and~\ref{sec:ach} have two further consequences, which correspond to the two extreme uses of a tape. An incompressible tape is useless as a fuel: no finite--state engine can extract work from it. And erasing a tape, which is the reverse of extracting work from it, costs work, and we characterize the minimal cost. Together, these two results are the individual--sequence counterparts of the familiar facts that a fully random tape has no fuel value and that erasing it costs $kT$ times its entropy.

\subsection{Incompressible sequences}

For the LZ complexity of the whole sequence (without restarts), we only have the one--sided relation $\bar H(x)\ge\limsup_N\rho_{\rm LZ}(x^N)$ of Section~\ref{sec:prelim}. This relation is enough to show that sequences that are incompressible by LZ78 are useless as fuel for any finite--state engine, even one designed with full knowledge of the sequence. Since the LZ complexity is computable, this gives a computable criterion.

\begin{Theorem}[Incompressible sequences]\label{thm:incomp}
If $\limsup_N\rho_{\rm LZ}(x^N)=\ln|\X|$, then $C(x)=0$. If, moreover, $\liminf_N\rho_{\rm LZ}(x^N)=\ln|\X|$, then $\limsup_N\sup_{{\cal E}(m,s)}\langle W_N\rangle/(NkT)\le0$ for every $m$ and $s$: for every $\varepsilon>0$ and all large $N$, no engine in ${\cal E}(m,s)$ extracts more than $\varepsilon kT$ per symbol, even if designed with full knowledge of $x^N$.
\end{Theorem}

\begin{proof}
The first statement follows from $\ln|\X|\ge\bar H(x)\ge\limsup_N\rho_{\rm LZ}(x^N)$ and Theorem~\ref{thm:cap}. For the second, fix a multiple $\ell$ of $m$, let $N$ be a multiple of $m$, and let $N'=\ell\lfloor N/\ell\rfloor$. The work on the first $N'$ symbols is bounded by \eqref{eq:convLZ} with $N'$ in place of $N$, and the remaining $N-N'<\ell$ symbols contribute at most $kT[(N-N')\ln|\X|+\ln s]$, by \eqref{eq:block} applied to each stage. Hence
\begin{equation}
\frac{\langle W_N\rangle}{NkT}\le\frac{N'}{N}\Bigl[\ln|\X|-\rho_{\rm LZ}(x^{N'})+\frac{\ln s}{\ell}+\delta(N',\ell)\Bigr]+\frac{\ell\ln|\X|+\ln s}{N}.
\end{equation}
Since $\liminf_N\rho_{\rm LZ}(x^{N'})\ge\liminf_N\rho_{\rm LZ}(x^N)=\ln|\X|$ and $\lim_N\delta(N,\ell)=(\ln2)/\ell$, the $\limsup_N$ of the right--hand side is at most $(\ln s+\ln2)/\ell$, and $\ell$ can be taken arbitrarily large.
\end{proof}

The second statement is stronger than the first, since $C(x)$ is defined with $\liminf_N$: the bound on the work per symbol holds, up to an arbitrarily small $\varepsilon$, for all large $N$, not only along a subsequence. Note also that incompressibility is a property of the sequence alone, whereas the theorem excludes work extraction by every finite--state engine.

\subsection{Erasure: a deterministic analogue of Landauer's principle}\label{sec:landauer}

Landauer's principle \cite{Landauer61,Bennett82} states that erasing one bit of information, i.e., resetting a random binary memory cell to a fixed state, costs at least $kT\ln2$ of work; more generally, erasing a random tape costs at least $kT$ times its entropy, and this cost can be approached by quasi--static operation. For a \emph{deterministic} tape, the Shannon entropy is zero, and the principle, in this form, says nothing: a device that knows the content of the tape can, in principle, reset it reversibly at no cost. Zurek \cite{Zurek89} and Bennett \cite{Bennett82} pointed out that, for a device that is a universal computer, the relevant quantity becomes the algorithmic (Kolmogorov) complexity of the content; see also \cite{KW20} for the thermodynamic costs of Turing machines, including the erasure of a long string. We now show that for finite--state engines, the relevant quantity is $\bar H(x)$, the same quantity that governs work extraction, so that the work needed to erase $x$ and the work that can be extracted from it are complementary.

Let $0\in\X$ and $0<\epsilon<1-1/|\X|$. An engine \emph{$\epsilon$--erases} $x^N$ if every symbol of its final output differs from $0$ with probability at most $\epsilon$. Define the finite--state erasure cost of $x$ as
\begin{equation}
\Gamma(x)=\lim_{\epsilon\to0}\ \inf_{m,s}\ \limsup_{N\to\infty}\ \inf\Bigl\{-\frac{\langle W_N\rangle}{NkT}:\ \text{engines in }{\cal E}(m,s)\text{ that $\epsilon$--erase }x^N\Bigr\},
\end{equation}
where $N$ ranges over multiples of $m$, and the limit exists since the quantity inside it is monotone in $\epsilon$. (The stricter criterion that every output segment differ from $0^m$ with probability at most $\epsilon$ gives the same $\Gamma(x)$: the lower bound below still applies, and the engines of the upper bound satisfy it.)

\begin{Theorem}[Erasure]\label{thm:erase}
For every infinite sequence $x$, $\Gamma(x)=\bar H(x)$. The upper bound is attained by engines that do not depend on $x$.
\end{Theorem}

Thus, the minimal work per symbol that finite--state engines need in order to erase $x$ is $kT\bar H(x)$: the Shannon entropy in Landauer's principle is replaced by the asymptotic normalized aligned block entropy of the individual sequence, which, by Proposition~\ref{prop:lz}, is also its block LZ complexity. A periodic sequence can be erased at a vanishing cost per symbol, whereas a sequence that is incompressible by finite--state machines costs $kT\ln|\X|$ per symbol, exactly as a fully random tape does, although it is deterministic. Moreover, by Theorem~\ref{thm:cap},
\begin{equation}
C(x)+\Gamma(x)=\ln|\X|,
\end{equation}
i.e., the fuel value of a tape and its erasure cost add up to $kT\ln|\X|$ per symbol, the fuel value of a blank tape, as in Bennett's analysis of the fuel value of a tape \cite{Bennett82}. As with work extraction, the optimal erasure cost is attained by engines that do not depend on $x$; they operate as a quasi--static process in space, as in Section~\ref{sec:ach}, but now towards a final landscape concentrated on the all--zero segment.

\begin{proof}
Let $\varepsilon_h=h(\epsilon)+\epsilon\ln(|\X|-1)$, the maximal entropy of a symbol that differs from $0$ with probability at most $\epsilon$. For the lower bound, repeat the derivation of Theorem~\ref{thm:conv}: for an $\epsilon$--erasing engine, each symbol of $\tilde Y_L$ differs from $0$ with probability at most $\epsilon$, so $H(\tilde Y_L)\le\ell\varepsilon_h$ replaces $H(\tilde Y_L)\le\ell\ln|\X|$. This gives $-\langle W_N\rangle/(NkT)\ge\hat H_\ell(x^N)/\ell-\varepsilon_h-(\ln s)/\ell$ for every multiple $\ell$ of $m$ and every multiple $N$ of $\ell$. Since a $\limsup_N$ over the multiples of $m$ is at least the $\limsup_N$ over the multiples of $\ell$, it follows, as in the proof of Theorem~\ref{thm:cap}, that $\Gamma(x)\ge\bar H(x)$.

For the upper bound, we use the cascade of Section~\ref{sec:ach} with a final landscape $R$ other than the uniform one: replace $Q^{1-\theta}$ in \eqref{eq:path} by $Q^{1-\theta}R^\theta$, so that $\Pi_1=R$. The derivation of Lemma~\ref{lem:landscape} goes through verbatim (H\"older's inequality now gives $\sum_bQ(b)^{1-\theta}R(b)^\theta\le1$), and gives \eqref{eq:landscape} with $m\ln|\X|$ replaced by $H(R)$ in the first term and by $\ln(1/\min_bR(b))$ in the last, and with $\Lambda_m=\max_b|\ln[R(b)/Q(b)]|$. Now fix $p$, use the standard choices \eqref{eq:standard}, i.e., $q=|\X|^{2p}$, $m=qp$ and $L=m^3$, let $A=|\X|^p$, and take the type--class code $Q_p$ as the initial landscape and $R=(1-\epsilon/2)\delta_{0^m}+(\epsilon/2)\cdot$uniform as the final one, where $\delta_{0^m}$ is the point mass at the all--zero segment. Under $R$, each symbol differs from $0$ with probability less than $\epsilon/2$, and choosing $\tau$ by Lemma~\ref{lem:finitetau} with $\gamma=\min\{1/m^2,\epsilon/2\}$, each final output segment is within total variation $\epsilon/2$ of $R$; so the engine $\epsilon$--erases $x^N$ for every $x$ and $N$. By the general version of Lemma~\ref{lem:landscape}, with $H(R)\ge0$, $\min_bR(b)\ge\epsilon/(2|\X|^m)$, and $\Lambda_m\le\Lambda'_m=m\ln|\X|+(A-1)\ln(q+1)+\ln(2/\epsilon)$,
\begin{equation}
-\frac{\langle W_N\rangle}{NkT}\le\frac{\hat H_p(x^N)}{p}+\frac{(A-1)\ln(q+1)}{m}+\frac{m\ln|\X|+\ln(2/\epsilon)+(\Lambda'_m)^2/2}{m^4}+\frac1{m^2}.
\end{equation}
As in \eqref{eq:schedule}, the second term on the right--hand side is at most $(2\ln|\X|+\ln2)|\X|^{-p}$, and $\Lambda'_m\le2m\ln|\X|+\ln(2/\epsilon)$, so that, for fixed $\epsilon$, the third and fourth terms are $O(1/m^2)$. Taking $\limsup_N$ and letting $p\to\infty$ along $p=k!$, along which $\bar H_p(x)\to\bar H(x)$ and $m\to\infty$, gives $\Gamma(x)\le\bar H(x)$.
\end{proof}

\section{Discussion}\label{sec:disc}

\paragraph{Segment length versus memory.} In the achievability part, the structure of $x$ is exploited entirely through the segment length, with no memory. In the converse, the segment length enters only through the requirement that $\ell$ be a multiple of $m$, while the memory enters through the penalty $(\ln s)/\ell$. So the converse does not exclude that an engine with a \emph{fixed} segment length exploits structure beyond its segments through memory, e.g., a one--cell ratchet that remembers previous symbols. Whether this is possible for autonomous engines with fixed energies is open; it requires copying information from the tape into the memory, which appears to be costly under detailed balance.

\paragraph{Time--dependent protocols.} If the energy levels may vary in time within each cycle, any stochastic map can be implemented with arbitrarily small entropy production \cite{OKW19}, and one may expect the capacity to be attainable with $m=1$, using memory. Such protocols, however, require an external agent that exchanges work with the device on a prescribed schedule, and so they are outside the autonomous model considered here.

\paragraph{Finite segment lengths.} Theorems~\ref{thm:ach} and~\ref{thm:univ} are asymptotic. We computed the work exactly at $\tau=\infty$ for binary sequences of length $N=2^{22}$. For an i.i.d.\ sample with $P(x_t=1)=0.2$, the target $\ln2-\hat H_1(x^N)$ is $0.1924\,kT$ per symbol, and the universal engine ${\cal U}[Q_1,4m,\infty]$ extracts $0.131$, $0.183$, and $0.1915\,kT$ per symbol for $m=16$, $256$, and $4096$, respectively. For the Thue--Morse sequence, whose structure is visible only with $p\ge2$, the target with $p=2$ is $(\ln2)/2\approx0.347\,kT$ per symbol, and ${\cal U}[Q_2,4m,\infty]$ extracts $0.107$ and $0.307\,kT$ per symbol for $m=16$ and $256$. The deficit decreases roughly like $[(|\X|^p-1)\ln q]/m$, in accordance with \eqref{eq:ach}, so segments of hundreds to thousands of cells are needed even for $p=1,2$. On a fair--coin sample (zero target), the universal engine consumes $0.044\,kT$ per symbol for $m=16$ and $0.0008\,kT$ for $m=4096$. By contrast, the engine designed for $x$ (end of Section~\ref{sec:types}), which has no redundancy, comes within $1\%$ of its target with $L=1000$ stages already for $m=p=4,8$, whereas a single stage extracts nothing, since its landscape is flat. Thus, the cost of universality is paid in the segment length, not in the number of stages. For the optimal engine of Theorem~\ref{thm:redund} (binary segments, $p=1$, $\tau=\infty$, $L=10^4$), the worst--case shortfall per segment is $1.1691$ nats for $m=4$ and $1.4461$ nats for $m=8$, compared with $\ln S_4(2)=1.1690$ and $\ln S_8(2)=1.4457$.

\paragraph{Work per symbol versus power.} All results concern the work per symbol in the limit of complete relaxation. The engines that attain the capacity have long segments, many stages and, by Lemma~\ref{lem:finitetau}, interaction times that must grow with $m$, since the relaxation time of a process on $|\X|^m$ states can grow with $m$. Thus, the power (work per unit time) of these engines may vanish, and the trade--off between work per symbol and power, as in finite--time thermodynamics, is not addressed here.

\paragraph{Open problems.} For memoryless engines fixed in advance, the optimal redundancy is $[\ln S_q(A)]/m$ per symbol (Theorem~\ref{thm:redund}). Can memory reduce it, e.g., by implementing a sequential universal code across segments, whose regret over the whole input grows only logarithmically in $N$? The Kraft inequality for work \eqref{eq:kraftwork} does not exclude this. Does the shortfall of Corollary~\ref{cor:most} hold for most inputs in general, not only for most periodic inputs? Can some single engine in a fixed class ${\cal E}(m,s)$ attain $C(x)$ for every $x$? We expect not.

\appendix
\renewcommand{\theLemma}{\thesection.\arabic{Lemma}}
\setcounter{Lemma}{0}
\section{Empirical Block Entropies}\label{app:blocks}

\begin{Lemma}\label{lem:mult}
For every infinite sequence $x$ and all positive integers $\ell$ and $k$, $\bar H_{k\ell}(x)\le \bar H_\ell(x)$.
\end{Lemma}
\begin{proof}
Let $N$ be a multiple of $k\ell$. Each aligned $k\ell$--block of $x^N$ is the concatenation of $k$ consecutive aligned $\ell$--blocks. Let $\tilde B=(\tilde C_1,\ldots,\tilde C_k)$ be a randomly chosen aligned $k\ell$--block, with $\tilde C_j\in\X^\ell$ its $j$--th $\ell$--block, so that $H(\tilde B)=\hat H_{k\ell}(x^N)$ by \eqref{eq:emp}. Every aligned $\ell$--block of $x^N$ appears as $\tilde C_j$ for exactly one aligned $k\ell$--block and one $j$, and so the average of the distributions of $\tilde C_1,\ldots,\tilde C_k$ is $\hat P_\ell$. By subadditivity and concavity of the entropy,
\begin{equation}
\hat H_{k\ell}(x^N)=H(\tilde C_1,\ldots,\tilde C_k)\le\sum_{j=1}^kH(\tilde C_j)\le kH(\hat P_\ell)=k\hat H_\ell(x^N).
\end{equation}
Dividing by $k\ell$ and taking $\limsup_N$ over the multiples of $k\ell$, which are among the multiples of $\ell$, gives $\bar H_{k\ell}(x)\le \bar H_\ell(x)$.
\end{proof}

\section{An Exponential--Family Bound}\label{app:lemmas}

\begin{Lemma}\label{lem:expfam}
Let $Q$ and $R$ be distributions on a finite set with full support, let $\Pi_\theta\propto Q^{1-\theta}R^\theta$, $0\le\theta\le1$, and let $\Lambda_m=\max_b|\ln[R(b)/Q(b)]|$. Then $D(\Pi_\theta\|\Pi_{\theta'})\le\frac12(\theta'-\theta)^2\Lambda_m^2$ for $0\le\theta<\theta'\le1$.
\end{Lemma}
\begin{proof}
Let $\phi=\ln(R/Q)$ and $\psi(\theta)=\ln\sum_bQ(b)e^{\theta\phi(b)}$, so that $\Pi_\theta(b)=Q(b)e^{\theta\phi(b)-\psi(\theta)}$. Then $\psi'(\theta)=\sum_b\Pi_\theta(b)\phi(b)$, and $\psi''(\theta)$ is the variance of $\phi$ under $\Pi_\theta$, so $0\le\psi''\le\Lambda_m^2$. Hence
\begin{equation}
D(\Pi_\theta\|\Pi_{\theta'})=\psi(\theta')-\psi(\theta)-(\theta'-\theta)\psi'(\theta)=\int_\theta^{\theta'}(\theta'-u)\psi''(u)\,du\le\tfrac12(\theta'-\theta)^2\Lambda_m^2.
\end{equation}
(For $R$ uniform on $\X^m$, $\Lambda_m$ is as in \eqref{eq:Lambda}.)
\end{proof}

\section{Markov Jump Processes}\label{app:markov}

Let $\{Z_t\}$ be a time--homogeneous Markov jump process on a finite set ${\cal Z}$ with transition rates $M[z\to z']\ge0$, $z\ne z'$, and transition matrices $P_t(z,z')=\Pr\{Z_t=z'|Z_0=z\}$. The process (equivalently, its rates) is called \emph{irreducible} if every state can be reached from every other state through jumps of positive rate, i.e., for all $z\ne z'$ there are states $z=z_0,z_1,\ldots,z_k=z'$ with $M[z_{i-1}\to z_i]>0$, $i=1,\ldots,k$. Write $\mu P_t$ for the distribution at time $t$ when $Z_0\sim\mu$.

\begin{Lemma}\label{lem:revP}
If the rates satisfy detailed balance with respect to a distribution $\Pi$ with full support, i.e., $\Pi(z)M[z\to z']=\Pi(z')M[z'\to z]$ for all $z\ne z'$, then $\Pi(z)P_t(z,z')=\Pi(z')P_t(z',z)$ for all $z,z'$ and all $t\ge0$.
\end{Lemma}
\begin{proof}
Let $G$ be the generator, i.e., $G(z,z')=M[z\to z']$ for $z\ne z'$, and $G(z,z)=-\sum_{z'\ne z}M[z\to z']$, and let $D$ be the diagonal matrix with entries $\Pi(z)$. Detailed balance says that $DG$ is symmetric, i.e., $DG=G^{\rm T}D$, hence $DG^k=(G^{\rm T})^kD$ for all $k$, and so $DP_t=De^{tG}=e^{tG^{\rm T}}D=P_t^{\rm T}D$, which is the assertion.
\end{proof}

\begin{Lemma}\label{lem:dp}
If $\tilde\Pi$ is a stationary distribution ($\tilde\Pi P_t=\tilde\Pi$ for all $t$) with full support, then $D(\mu P_t\|\tilde\Pi)\le D(\mu\|\tilde\Pi)$ for every distribution $\mu$ and every $t\ge0$.
\end{Lemma}
\begin{proof}
$\mu P_t$ and $\tilde\Pi=\tilde\Pi P_t$ are the outputs of the same channel $P_t$ with inputs $\mu$ and $\tilde\Pi$. By the data processing inequality for divergence \cite{CT06}, $D(\mu P_t\|\tilde\Pi P_t)\le D(\mu\|\tilde\Pi)$.
\end{proof}

\begin{Lemma}\label{lem:conv}
If the process is irreducible, then there is a unique stationary distribution $\Pi$, and $\max_{z,z'}|P_t(z,z')-\Pi(z')|\to0$ as $t\to\infty$.
\end{Lemma}
\begin{proof}
For an irreducible process on a finite set, $P_t(z,z')>0$ for all $z,z'$ and all $t>0$. Hence $P_1$ is a stochastic matrix with all entries positive, and $P_{t}=P_1^{\lfloor t\rfloor}P_{t-\lfloor t\rfloor}$. By the convergence theorem for positive stochastic matrices, $P_1^j$ converges geometrically fast to the matrix $\bar\Pi$ all of whose rows are equal to $\Pi$, where $\Pi$ is the unique stationary distribution of $P_1$, which is also stationary for all $P_t$ (since $P_t$ and $P_1$ commute, $\Pi P_t$ is stationary for $P_1$, and so $\Pi P_t=\Pi$ by uniqueness). Since $P_{t-\lfloor t\rfloor}$ is stochastic and $\Pi$ is stationary, $P_t\to\bar\Pi$ as well. (See, e.g., \cite{Norris97}.)
\end{proof}

\begin{Lemma}\label{lem:finitetau}
Consider the engine of Section~\ref{sec:ach} with initial landscape $Q$, final landscape $R$ (uniform, or general as in Section~\ref{sec:landauer}), $L$ stages and interaction time $\tau$; its stage $r$ has canonical distribution $\Pi_{\theta_r}$, $\theta_r=r/L$. For every $\gamma>0$ there is $\tau_0$, depending only on $Q$, $R$, $L$ and the rates (not on $x$ or $N$), such that for all $\tau\ge\tau_0$, every $x$ and every $N$ that is a multiple of $m$: (i) the expected work on $x^N$ is at least its value at $\tau=\infty$ minus $\gamma NkT$; (ii) every final output segment has a distribution within total variation $\gamma$ of $R$.
\end{Lemma}

\begin{proof}
\emph{Step 1: closeness of the kernels.} Let $\Phi_{r,\tau}(b|a)$ be the one--cycle kernel of stage $r$ (the probability that a segment entering with content $a$ leaves with content $b$), and let
\begin{equation}
\varepsilon(\tau)=\max_{1\le r\le L}\ \max_a\sum_b\bigl|\Phi_{r,\tau}(b|a)-\Pi_{\theta_r}(b)\bigr|.
\end{equation}
Since the Markov jump process of each stage is irreducible, $\varepsilon(\tau)\to0$ as $\tau\to\infty$ (Lemma~\ref{lem:conv}). Note that $\varepsilon(\tau)$ depends only on the engine, not on $x$ or $N$.

\emph{Step 2: the input of a stage on one segment.} Since the stages are memoryless, the output of stage $r$ on segment $i$ depends only on the input of stage $r$ on segment $i$. Let $\mu$ be the distribution of that input at the given $\tau$, and $\mu^\infty$ its distribution at $\tau=\infty$. For $r=1$, $\mu=\mu^\infty$ is the point mass at $\bx_i$. For $r\ge2$, $\mu=\sum_a\mu'(a)\Phi_{r-1,\tau}(\cdot|a)$, where $\mu'$ is the distribution of the input of stage $r-1$ on segment $i$, while $\mu^\infty=\Pi_{\theta_{r-1}}$; hence $\sum_b|\mu(b)-\mu^\infty(b)|\le\varepsilon(\tau)$.

\emph{Step 3: the work of a stage on one segment.} Let $E_{\max}=\max_{r,b}|E_r(b)|$. The expected work of stage $r$ on segment $i$ is $\sum_a\mu(a)f_\tau(a)$, where $f_\tau(a)=\sum_b\Phi_{r,\tau}(b|a)E_r(b)-E_r(a)$, and at $\tau=\infty$ it is $\sum_a\mu^\infty(a)f_\infty(a)$, with $f_\infty(a)=\sum_b\Pi_{\theta_r}(b)E_r(b)-E_r(a)$. Since $|f_\tau(a)-f_\infty(a)|\le E_{\max}\varepsilon(\tau)$ and $|f_\infty(a)|\le2E_{\max}$,
\begin{align}
\Bigl|\sum_a\mu(a)f_\tau(a)-\sum_a\mu^\infty(a)f_\infty(a)\Bigr|&\le\sum_a\mu(a)|f_\tau(a)-f_\infty(a)|+\sum_a|\mu(a)-\mu^\infty(a)|\,|f_\infty(a)|\nonumber\\
&\le E_{\max}\varepsilon(\tau)+2E_{\max}\varepsilon(\tau)=3E_{\max}\varepsilon(\tau).
\end{align}

\emph{Step 4: conclusion.} Summing over the $L$ stages and the $J$ segments, the work at $\tau$ differs from the work at $\tau=\infty$ by at most $3E_{\max}LJ\varepsilon(\tau)$. The distribution of each final output segment is $\sum_a\mu(a)\Phi_{L,\tau}(\cdot|a)$ for some $\mu$, so its total variation distance from $\Pi_1=R$ is at most $\varepsilon(\tau)/2$. Choose $\tau_0$ such that $3E_{\max}L\varepsilon(\tau)\le\gamma mkT$ and $\varepsilon(\tau)\le2\gamma$ for all $\tau\ge\tau_0$. Then the deficit is at most $\gamma mJkT=\gamma NkT$, and the final segments are within total variation $\gamma$ of $R$, for every $x$ and $N$.
\end{proof}

\section{Kraft Inequality for Work with Memory}\label{app:kraftmem}

\begin{Lemma}\label{lem:kraftmem}
For every engine in ${\cal E}(m,s)$, every initial distribution of its ratchet states (independent of the input), and every $N$ that is a multiple of $m$, inequality \eqref{eq:kraftwork} holds.
\end{Lemma}

\begin{proof}
Let the engine consist of $L$ stages, where stage $r$ has the set ${\cal S}_r$ of $s_r$ ratchet states, energy levels $E_r(y,\sigma)$, $(y,\sigma)\in\X^m\times{\cal S}_r$, and transition matrix $P_r$ of its Markov jump process over one interaction interval, and let $J=N/m$.

\emph{Global configurations.} A global configuration is a vector $\zeta=(y_1,\ldots,y_J,\sigma_1,\ldots,\sigma_L)$, where $y_i\in\X^m$ is the current content of segment $i$, i.e., its content on the tape of the last stage that has processed it (initially, the input segment $\bx_{i-1}$), and $\sigma_r\in{\cal S}_r$ is the current ratchet state of stage $r$. Let ${\cal G}$ be the set of global configurations; then $|{\cal G}|=|\X|^Ns_1\cdots s_L$.

\emph{Cycles as transitions.} Each cycle of the engine is a pair $(i,r)$: stage $r$ processing segment $i$. We list the $JL$ cycles in the order $(1,1),(1,2),\ldots,(1,L),(2,1),\ldots,(J,L)$, i.e., segment by segment. This order is compatible with the cascade: when cycle $(i,r)$ starts, segment $i$ has been processed by stages $1,\ldots,r-1$, and stage $r$ has processed segments $1,\ldots,i-1$, so that its ratchet state is the one left by cycle $(i-1,r)$. Since each cycle depends only on its own input segment and ratchet state, the joint distribution of all the Markov--state trajectories, and hence of the work, is the same as under any other compatible timing. Cycle $c=(i,r)$ changes only the components $(y_i,\sigma_r)$ of the global configuration, which form the Markov state $z=(y_i,\sigma_r)$ of stage $r$; let
\begin{equation}
K_c(\zeta,\zeta')=\left\{\begin{array}{ll}P_r(z,z') & \text{if $\zeta$ and $\zeta'$ agree in all components other than $(y_i,\sigma_r)$,}\\ 0 & \text{otherwise,}\end{array}\right.
\end{equation}
where $z$ and $z'$ are the Markov states of stage $r$ contained in $\zeta$ and $\zeta'$, and let $E_c(\zeta)=E_r(z)$. Let $\zeta_0$ be the initial global configuration and $\zeta_c$ the global configuration after cycle $c$. Then $\zeta_0,\zeta_1,\ldots,\zeta_{JL}$ is a Markov chain with transition matrices $K_1,\ldots,K_{JL}$, and, since the work of a jump is the corresponding energy difference, the work of cycle $c$ is $E_c(\zeta_c)-E_c(\zeta_{c-1})$, so that $W_N=\sum_{c=1}^{JL}[E_c(\zeta_c)-E_c(\zeta_{c-1})]$.

\emph{Detailed balance.} By Lemma~\ref{lem:revP}, $e^{-\beta E_r(z)}P_r(z,z')=e^{-\beta E_r(z')}P_r(z',z)$, and hence, for every $c$ and all $\zeta,\zeta'\in{\cal G}$,
\begin{equation}
K_c(\zeta,\zeta')e^{\beta[E_c(\zeta')-E_c(\zeta)]}=K_c(\zeta',\zeta),
\end{equation}
since both sides vanish unless $\zeta$ and $\zeta'$ agree outside $(y_i,\sigma_r)$.

\emph{The identity.} For every initial configuration $\zeta_0$,
\begin{equation}
\bigl\langle e^{\beta W_N}\bigr\rangle_{\zeta_0}=\sum_{\zeta_1,\ldots,\zeta_{JL}}\prod_{c=1}^{JL}K_c(\zeta_{c-1},\zeta_c)e^{\beta[E_c(\zeta_c)-E_c(\zeta_{c-1})]}=\sum_{\zeta_1,\ldots,\zeta_{JL}}\prod_{c=1}^{JL}K_c(\zeta_c,\zeta_{c-1}).
\end{equation}
Summing over $\zeta_0$, then over $\zeta_1$, and so on up to $\zeta_{JL-1}$, each sum $\sum_{\zeta_{c-1}}K_c(\zeta_c,\zeta_{c-1})$ equals $1$, and the remaining sum over $\zeta_{JL}$ gives $|{\cal G}|$. Thus,
\begin{equation}
\sum_{x^N\in\X^N}\ \sum_{\sigma\in{\cal S}_1\times\cdots\times{\cal S}_L}\bigl\langle e^{\beta W_N}\bigr\rangle_{x^N,\sigma}=|\X|^Ns_1\cdots s_L,
\end{equation}
where $\sigma$ is the vector of initial ratchet states.

\emph{Conclusion.} Since all terms are nonnegative, for every fixed $\sigma$, $\sum_{x^N}\langle e^{\beta W_N}\rangle_{x^N,\sigma}\le|\X|^Ns_1\cdots s_L\le|\X|^Ns$. If the initial ratchet states are drawn from a distribution $\pi$, independently of the input, then $\langle e^{\beta W_N}\rangle_{x^N}=\sum_\sigma\pi(\sigma)\langle e^{\beta W_N}\rangle_{x^N,\sigma}$, and so, again, $\sum_{x^N}\langle e^{\beta W_N}\rangle_{x^N}\le|\X|^Ns$. Finally, by Jensen's inequality, $e^{\beta\langle W_N(x^N)\rangle}\le\langle e^{\beta W_N}\rangle_{x^N}$, and dividing by $|\X|^N=e^{N\ln|\X|}$ gives \eqref{eq:kraftwork}.
\end{proof}

\end{document}